\documentclass{ieeeaccess}
\usepackage{cite}
\usepackage{amsmath,amssymb,amsfonts}
\usepackage{algorithmic}
\usepackage{graphicx}
\usepackage{textcomp}
\def\BibTeX{{\rm B\kern-.05em{\sc i\kern-.025em b}\kern-.08em
    T\kern-.1667em\lower.7ex\hbox{E}\kern-.125emX}}

\usepackage{url, pifont}
\newcommand*\rot{\rotatebox{90}}
\newcommand*\OK{\ding{51}}

\begin{document}


\title{Blockchain as privacy and security solution for smart environments: A Survey}

\author{
\uppercase{Maad Ebrahim}\authorrefmark{1},
\uppercase{Abdelhakim Hafid}\authorrefmark{1} \IEEEmembership{Member, IEEE}, and
\uppercase{Etienne Elie}\authorrefmark{2}
}
\address[1]{Department of Computer Science and Operations Research (DIRO),
University of Montreal, Montreal, QC H3T1J4 Canada}
\address[2]{Intel Corporation, 2200 Mission College Blvd, Santa Clara, CA 95054}


\markboth
{Ebrahim \headeretal: Blockchain as privacy and security solution for smart environments: A Survey}
{Ebrahim \headeretal: Blockchain as privacy and security solution for smart environments: A Survey}

\corresp{Corresponding author: Maad Ebrahim (e-mail: maad.ebrahim@umontreal.ca).}

\begin{abstract}
Blockchain was always associated with Bitcoin, cryptocurrencies, and digital asset trading. However, the benefits of Blockchain are far beyond that. It has been recently used to support and augment many other technologies, including the Internet-of-Things (IoT). IoT, with the help of Blockchain, paves the way for futuristic smart environments, like smart homes, smart transportation, smart energy trading, smart industries, smart supply chains, and more. To enable these smart environments, IoT devices, machines, appliances, and vehicles, will need to intercommunicate without the need for a centralized trusted party. Blockchain can replace third trusted parties by providing secure means of decentralization in such trustless environments. They also provide security enforcement, privacy assurance, authentication, and other important features to IoT ecosystems. Besides the benefits of Blockchain-IoT integration for smart environments, other technologies also have important features and benefits that attracted the research community. Software-Defined Networking (SDN), Fog, Edge, and Cloud Computing technologies, for example, play an important role in enabling realistic IoT applications. Moreover, the integration of Machine Learning and Artificial Intelligence (AI) algorithms provides smart, dynamic, and autonomous decision-making capabilities for IoT devices in smart environments. To push the research further in this domain, we provide in this paper a comprehensive survey that includes state-of-the-art technological integration, challenges, and solutions for smart environments, and the role of Blockchain and IoT technologies as the building blocks of such smart environments. We also demonstrate how the level of integration between these technologies has increased over the years, which brings us closer to the futuristic view of smart environments. We further discuss the current need to provide general-purpose Blockchain platforms that can adapt to different design requirements of different applications and solutions. Finally, we provide a simplified architecture of futuristic smart environments that integrates all these technologies, showing the advantage of such integration.
\end{abstract}

\begin{keywords}
Artificial Intelligence (AI), Blockchain, Cloud Computing, Edge Computing, Fog Computing, Internet-of-Things (IoT), Software-Defined Networking (SDN), Smart Environments
\end{keywords}

\titlepgskip=-15pt
\maketitle

\section{Introduction}
\label{sec:introduction}

Technology development is progressing rapidly, even faster than the expectations decades ago. The reason for such explosion in the technology is the huge effort that is being conducted by the research community, which aims to facilitate human life via developing a futuristic view of a smarter earth. There has been a lot of academic work and industrial adoption to create and implement prototypes of smart cities, which include smart homes, smart factories, smart cars, smart transportation, and various smart human gadgets. A lot of core technologies helped reaching this point of success for this futuristic human civilization. These technologies include, but not limited to, Internet-of-Things (IoT), Software-Defined Networking (SDN), Artificial Intelligence (AI), and Cloud, Fog, and Edge Computing. In addition, Blockchain was able to augment those technologies with more features that are essential for the full automation that is needed in smart environments. The role of IoT and Blockchain in Smart environments can be understood from the increase in popularity of the web search terms that are shown in Fig. \ref{fig:trends} over the last few years.

\Figure[t!](topskip=0pt, botskip=0pt, midskip=0pt)[width=0.48\textwidth]{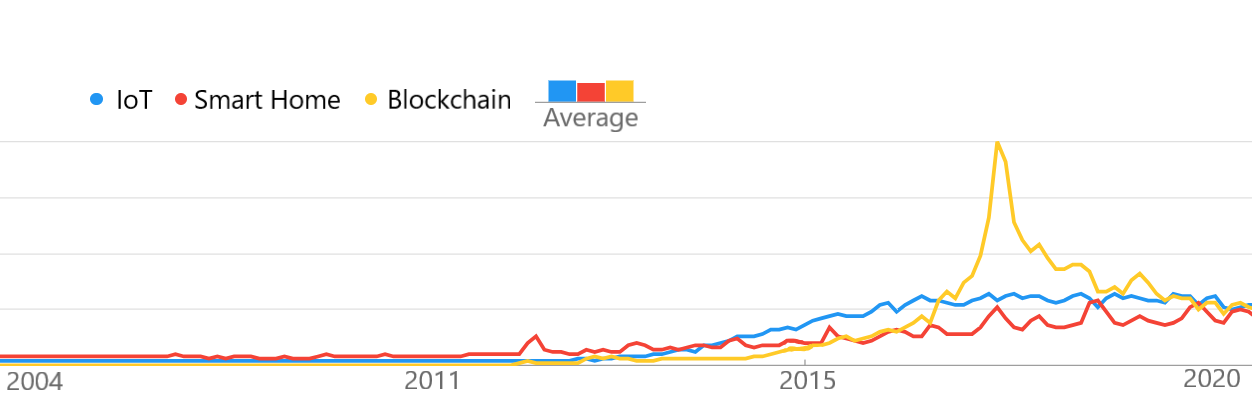}
{Google Trends search interest for IoT, Smart Home and Blockchain terms between 2004-2020.\label{fig:trends}}

IoT enabled every physical device to be connected to the internet in order to communicate with other physical devices and services. This can enable, for example, a fridge in the future to automatically detect missing items for its users and automatically order those items from the nearest grocery store. The system in the grocery store can respond to this order and automatically receive the required monetary value when accepting the transaction. The grocery items are then automatically collected and sent to the customer using a self driving vehicle. In this futuristic world, the owner of the house does not need to set his smart alarm clock, as it is automatically synchronized to wake him up for his next meeting. To reach his meeting on time, his self-driving car chooses the fastest and safest path using real-time information of the traffic in the city. The car is notified in real-time of nearby accidents in order to optimize its route. This car can communicate with other smart vehicles along the route to provide the safest driving experience for all vehicles on the road.

Besides IoT, SDN enables for dynamic and programmable control and management of the underlying network in a smart way. SDN best suits IoT networks, since they change rapidly in terms of the number of devices, their locations, and the amount of data they send. Moreover, SDN enables for the integration of AI and machine learning into the decision-making process of load balancing, computational offloading, traffic control, and data flow in the network. SDN is considered one of the major factors in enabling IoT, and hence smart cities innovation \cite{ghosh2020sdniotbased}. However, it requires intensive computations and storage requirements that cannot be provided by the low-power and limited-resources IoT devices. That is why Cloud, Fog, and Edge Computing technologies were introduced to basically help providing storage and computation resources as paid services to manage such networks.

Cloud Computing provides theoretically unlimited storage and computation resources as paid services. They are usually located in central data centers located in distant geographical locations. The distance can burden the core network by the huge amount of traffic created by IoT devices. This distance also increases the service response time (delay) for IoT devices, specially when they need a feedback on their requests. Such delay might not be acceptable for time-sensitive IoT applications, such as self-driving cars, where a delay in milliseconds can cause catastrophic incidents. Hence, technologies such as Fog and Edge Computing provide solutions to these problems by bringing those resources closer to the IoT infrastructure. They minimize the delay and save the network bandwidth by performing preliminary pre-processing and analysis on IoT data before sending it to the cloud for heavier processing and permanent storage.

There is one missing connection for all those technologies in order to enable the futuristic concept of "trustless" smart cities we talked about earlier. To enable the communication among multiple IoT devices that are usually manufactured/owned by different organizations, a third trusted party is usually needed to provide the trust among the devices performing the transactions. Blockchain can act here as that missing connection in order to provide this trust mechanism in a decentralized manner. Blockchain can also be used as a mechanism to permanently log the transactions executed by IoT devices, manage digital assets trading, and perform monetary transactions between them. Actually, Blockchains can do much more than that; they can support and enrich the development of the IoT industry, and they can mitigate many of the current limitations in SDN, Cloud, Fog, and Edge solutions for IoT applications.

In this paper, we present a comprehensive survey that shows what Blockchain can provide by its integration with other technologies to build the foundation for future smart environments. This integration is oriented towards IoT applications and supported by different emerging technologies (see Fig. \ref{fig:integration}). We also show few applications that are brought to life with the help of Blockchain and its integration with those technologies. We further discuss some of the challenges and open research problems of Blockchain and its integration with those technologies. These challenges need to be addressed by the research community in order to provide a ready-to-go Blockchain-based decentralization platform for smart environments.

\Figure[t!](topskip=0pt, botskip=0pt, midskip=0pt)[width=0.48\textwidth]{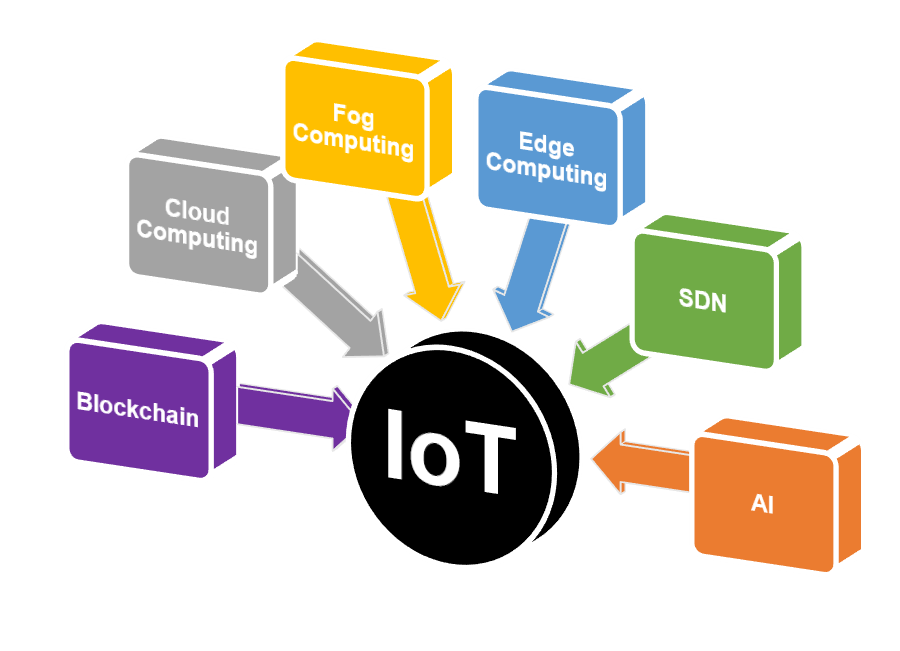}
{The integration of IoT with Blockchain, SDN, AI, Cloud, Fog, and Edge Computing technologies.\label{fig:integration}}

The rest of the paper is organized as follows. Section \ref{sec:compare} compares this work with existing surveys. We then start by introducing our definition for smart environments in Section \ref{sec:smartEnv}. Section \ref{sec:blockchain} briefly introduces Blockchain and some of its applications. In Section \ref{sec:IoT}, we present IoT-Blockchain integration and some applications for such integration. Section \ref{sec:Cloud} describes the benefits of integrating Cloud, Fog, and Edge Computing technologies into Blockchain-IoT ecosystems. Sections \ref{sec:SDN} and \ref{sec:AI} show how SDN and AI technologies, respectively, help supporting Blockchain-IoT infrastructures. In Section \ref{sec:res}, we discuss some challenges and open research problems that need to be addressed to successfully enable smoother technological integration. Finally, we present in Section \ref{sec:con} a general smart environment architecture that integrates the technologies presented in this survey to satisfy its requirements.

\section{Our Work and Existing Surveys}
\label{sec:compare}
Presenting Blockchain integration into multiple technologies in the context of smart environments is what makes this work unique compared to previous surveys. We show how this integration is able to augment those technologies, and how it helps pave the way for smart environments of the future. Beside giving a brief introduction about Blockchain and its applications in smart environments, this work discusses Blockchain integration with IoT to allow for full automation in smart devices. We then study how Blockchain-based IoT solutions are augmented with SDN, Cloud, Fog, and Edge Computing to enhance the capabilities of IoT applications. Finally, we study the impact of AI and machine learning algorithms to make such solutions even smarter. 

Table \ref{tab:compareSurv} shows the level of technological integration to Blockchain-IoT solutions in reviews and surveys. The table shows that most existing surveys do not consider the inclusion of all these technologies to enhance Blockchain-IoT integration. In addition, existing surveys do not elaborate on the direct impact such integration on smart environment applications. Most surveys focus on integrating Cloud Computing to mitigate the resource limitations of IoT devices while considering Fog and Edge technologies to provide privacy and minimize the delay. However, there is a lack of surveys covering the recent interest in using SDN and AI technologies for dynamic network management and complex optimization problems, respectively.

\begin{table}
  \caption{The integration of Cloud, Edge, Fog, SDN, and AI technologies in Blockchain-IoT solutions.}
  \label{tab:compareSurv}
  \setlength{\tabcolsep}{3pt}
    \begin{tabular}{|l|c| *{4}c |}
        \hline
        \textbf{Authors} & \textbf{Year} & \multicolumn{4}{c|}{\textbf{Integration with}} \\[1ex]
        & & \rot{Cloud} & \rot{Edge/Fog} & \rot{SDN} & \rot{AI} \\\hline\hline    

        Conoscenti \textit{et al.} \cite{systematic2016Conoscenti} & 2016 &&&& \\\hline
        Christidis and Devetsikiotis \cite{Christidis2016BusinessModel} & 2016 &&&& \\\hline
        Reyna \textit{et al.} \cite{REYNA2018173} & 2018 &\OK&&& \\\hline
        Ramachandran and Krishnamachar \cite{ramach2018blockchain} & 2018 &&&& \\\hline
        Panarello \textit{et al.} \cite{Integeration2018Penarello} & 2018 &\OK&&& \\\hline
        Fernández-Caramés and Fraga-Lamas \cite{Review2018} & 2018 &\OK&\OK&\OK& \\\hline
        Banerjee \textit{et al.} \cite{BANERJEE2018149} & 2018 &\OK&\OK&\OK&\OK \\\hline
        Atlam \textit{et al.} \cite{soton421529} & 2018 &&&& \\\hline
        Zheng \textit{et al.} \cite{zheng2018Blockchain} & 2018 &&&& \\\hline
        Ali \textit{et al.} \cite{Ali2019Comprehensive} & 2019 &\OK&\OK&\OK&\OK \\\hline
        Dai \textit{et al.} \cite{BCIoTSurvey} & 2019 &\OK&\OK&\OK&\OK \\\hline
        Ferrag \textit{et al.} \cite{8543246} & 2019 &\OK&\OK&\OK& \\\hline
        Makhdoom \textit{et al.} \cite{MAKHDOOM2019251} & 2019 &\OK&\OK&& \\\hline
        Salah \textit{et al.} \cite{Blockchain2019AI} & 2019 &\OK&\OK&&\OK \\\hline
        Yang \textit{et al.} \cite{8624417} & 2019 &\OK&\OK&\OK&\OK \\\hline
        Lao \textit{et al.} \cite{SurveyIoT2020} & 2020 &&\OK&& \\\hline
        Alharbi \cite{Alharbi2020SDNBlockchain} & 2020 &\OK&\OK&\OK& \\\hline
        LI \textit{et al.} \cite{Wenjuan2020BlockchainSDN} & 2020 &&&\OK& \\\hline
        Luo \textit{et al.} \cite{Luo2020} & 2020 &\OK&\OK&& \\\hline
        Xie \textit{et al.} \cite{BlockchainSurvey2020} & 2020 &\OK&&& \\\hline
        Mohanta \textit{et al.} \cite{MOHANTA2020100227} & 2020 &\OK&\OK&&\OK \\\hline
        Chamola \textit{et al.} \cite{9086010} & 2020 &\OK&&&\OK \\\hline
    \end{tabular}
\end{table}

Stojkoska and Trivodaliev \cite{RISTESKASTOJKOSKA20171454}, for example, focused on the role of IoT in smart home applications. In addition, Bhushan \textit{et al.} \cite{BHUSHAN2020102360} discussed the integration of Blockchain to support IoT applications in smart cities. Even though our work is oriented towards smart environments, we also study the effect of technological integration to establish such smart environments. We cover how other technologies help mitigate several problems in Blockchain-based IoT smart systems. Other surveys also discussed IoT integration with other technologies, like AI \cite{AI4IoT}, or SDN \cite{SDN2017Survey}. However, these surveys do not consider smart environment applications, and do not cover the benefits of Blockchain decentralization properties.

The majority of the surveys only focus on Blockchain and IoT integration, including the benefits and challenges of such integration \cite{soton421529, ramach2018blockchain, MAKHDOOM2019251, systematic2016Conoscenti, Integeration2018Penarello}. Ali \textit{et al.} \cite{Ali2019Comprehensive}, for example, reviewed Blockchain-based platforms and services that are used to augment IoT applications. Similarly, Lao \textit{et al.} \cite{SurveyIoT2020} covered the use of Blockchain to address IoT limitations and secure IoT networks. They also gave a comprehensive overview of IoT-Blockchain applications, including architectures, communication protocols, and traffic models for such applications. There are several other surveys that only focus on the effort to secure IoT networks using Blockchain \cite{BANERJEE2018149, BCIoTAccess}. Riabi \textit{et al.} \cite{BCIoTAccess}, for example, covered contributions that use Blockchain to mitigate single-points-of-failures in centralized access control architectures for IoT devices.

Even though the majority of the surveys only deal with Blockchain-IoT integration, some of them have different focus or interest. Reyna \textit{et al.} \cite{REYNA2018173} did cover Blockchain-IoT integration, in particular, running Blockchain on IoT devices.  Fernández-Caramés and Fraga-Lamas \cite{Review2018} reviewed Blockchain-IoT integration in healthcare, logistics, smart cities, and energy management systems. Likewise, Ferrag \textit{et al.} \cite{8543246} focused on Blockchain-IoT integration for applications in Internet-of-Vehicles (IoV), Internet-of-Energy (IoE), Internet-of-Cloud (IoC), and Edge Computing. In addition to those applications, Mohanta \textit{et al.} \cite{MOHANTA2020100227} did overview existing security solutions for IoT networks using Blockchain and AI technologies. Garcia \cite{Garcia2020} reviewed the integration of AI, IoT, and Blockchain from the taxation, legal, and economical point of views.

There are also other surveys that focus on the integration of Blockchain and other technologies outside the context of IoT. Ekramifard \textit{et al.} \cite{Ekramifard2020}, for example, produced a systematic literature review on the integration of Blockchain with AI; particularly identifying the applications that can benefit from such integration. Similarly, Salah \textit{et al.} \cite{Blockchain2019AI} surveyed the way Blockchain can enhance and solve AI limitations. They also reviewed the role of Blockchain in achieving decentralized AI schemes. Additionally, Akter \textit{et al.} \cite{Akter2020} investigated a diverse set of applications in the literature that are based on Blockchain, AI, and Cloud technologies. Xie \textit{et al.} \cite{BlockchainSurvey2020} surveyed Blockchain-based solutions to augment the Cloud Computing technology. They did study the role of Blockchain to provide decentralized Cloud Exchange services.

Other surveys focused on Blockchain integration with Fog and/or Edge computing technologies. For instance,  Baniata and Kertesz \cite{BCandFog} covered contributions that integrate Blockchain with Fog Computing. likewise, Yang \textit{et al.} \cite{8624417} provided a survey on Blockchain integration with Edge computing, which can help decentralize network management. Moreover, SDN integration with Blockchain has been also presented in other surveys \cite{Alharbi2020SDNBlockchain, Wenjuan2020BlockchainSDN}. Alharbi \cite{Alharbi2020SDNBlockchain}, for example, surveyed existing papers that secure SDN architectures from different attacks using Blockchain. Moreover, LI \textit{et al.} \cite{Wenjuan2020BlockchainSDN} reviewed how Blockchain and SDN technologies complement each other when integrated together.

Our work goes far beyond few technological integration between Blockchain and IoT to mitigate some of their limitations. We review existing work according to different levels of technological integration, which is oriented towards smart environment applications. We briefly discuss the benefits of Blockchain technology itself, and the benefits of Blockchain-IoT solutions in smart environments. Then, we present the benefits of integrating Cloud, Fog, Edge, SDN, and AI technologies in Blockchain-IoT smart systems. We then present open research problems to be addressed to create smooth technological integration for smart futuristic environments. Finally, we provide a simplified smart environment architecture that shows how Blockchain help integrating the various technologies discussed in this paper.

\section{Smart Environments}
\label{sec:smartEnv}

Over history, humans have always created innovative solutions to make their lives easier. The recent technological inventions allowed us to live in an environment that was considered science fiction in the past. However, scientists always wanted to push this further by creating a smarter world where automation is included in every aspect of human lives. This was only possible through introducing IoT technology, where IoT sensors and actuators are embedded in physical devices and machines, making them able to interconnect through the internet. IoT, with the help of Big data analysis, AI, Machine Learning, and many other innovative technologies allowed for the realization of these smart environments \cite{smartEnvironments1_2021, smartEnvironments2_2021}.

Governments, industries, and scientists are all racing towards creating and prototyping smart cities for boosting the life quality of their citizens. Smart homes, for example, provide a futuristic domestic environment that delivers a technologically advanced living experience for people \cite{smartCities2_2021}. While smart education includes smart campuses, smart universities, and smart classrooms for students in these smart cities \cite{smartEducation_2021}. Various innovative solutions were used to mitigate different challenges in realizing these environments, like using Blockchain to secure smart city applications \cite{smartCities1_2021}. The difficulty in realizing these applications increases as the human interactions with the smart infrastructure get more complex, as in the case of smart transportation systems \cite{smartCities3_2021}. Therefore, the research in these complex smart infrastructures, including smart transportation systems, had become a dominant research topic in the context of smart environments \cite{smartTransporation_2021}.

The fourth industrial revolution, also called Industry 4.0, is another important component of smart environments, which could only be realized after introducing smart factories and smart supply chain systems \cite{smartIndustry4_2021}. To decrease the cost while increasing the quality of mass production, businesses stood up for shifting from traditional manufacturing to smart factories \cite{smartIndustry1_2021}. The introduction of Smart Industry allows for the automation of intelligent predictive maintenance strategies, which provide major cost saving over time-based preventive maintenance \cite{smartIndustry2_2021}. The fourth industrial revolution in smart cities also led to the development of distributed smart energy trading systems, which require Blockchain for security and reliability \cite{smartIndustry3_2021}.

Smart farming is another component of smart environments, which was enabled by IoT, Wireless Sensor Networks (WSN), Cloud Computing, Fog Computing, as well as Big data analytics \cite{smartFarming3_2021}. Big data analytic plays an important role in bringing real-time decision-making capabilities into smart farming environments by obtaining valuable information from the collected data \cite{smartFarming1_2021}. For example, machine learning can automate decision-making in smart farming environments by predicting soil drought and crop productivity \cite{smartFarming2_2021}. Smart vehicles also exist in almost all smart environments, including smart farming, smart factories, smart cities, and smart transportation systems, and we also see how Blockchain can secure the transactions between these smart vehicles \cite{smartCars_2021}. 

IoT paved the way to pervasive computing, also called ubiquitous computing, which is the interconnection of sensors with computing capabilities through the internet. These smart devices are the main building blocks for smart environments, which aim at providing a comfortable living experience for humans via performing repetitive or risky tasks using these devices. Furthermore, other technologies are needed to get the full benefit out of these devices, including AI, machine learning, Big data analytics, as well as computer networks, parallel and distributed computing, and much more. The realization of smart environments needs a lot of work, so in this paper, we focus on how Blockchain provides security, privacy, and other features into smart environments with the help of other technologies. The outcomes of this research spot the light on what Blockchain can provide for smart environments. Besides, it proposes a simplified IoT-based smart environment architecture using Blockchain with Cloud/Fog computing, SDN, and AI technologies.

\section{Blockchain and its Applications} 
\label{sec:blockchain}

Starting by Bitcoin in 2008, Blockchain went far beyond the world of cryptocurrencies as it was proposed by Nakamoto \textit{et al.} \cite{nakamoto2008Bitcoin}. Blockchain enabled many smart environment applications and solutions as has been clearly seen in recent studies. For example, Blockchain provides authentication and authorization for smart city applications \cite{BC_Auth_2021}, like managing real estate deals in smart cities \cite{BC_FERL_2021}. Blockchain was also used as a core framework to secure smart vehicles \cite{BC_FERL_2021} and smart grid systems \cite{BC_Grids_2021}.

Blockchain is essentially a distributed ledger that is made up of blocks (list of transactions) that are backward-connected through hash pointers as shown in Fig. \ref{fig:blockchain}. Besides the transactions' Merkel tree inside each block. These pointers guarantee the immutability of this ledger. Different consensus algorithms, such as Proof-of-Work (PoW) \cite{nakamoto2008Bitcoin}, have been proposed to securely update the ledger without the need to a centralized entity. In PoW, the nodes in the network compete to solve a puzzle that is solved only by trying different Nonce values. Blockchain can provide pseudonymity and traceability, which are very useful in dozens of domains. Those features are achieved with the help of different technologies, including cryptography, hashing, and digital signatures.

\Figure[t!](topskip=0pt, botskip=0pt, midskip=0pt)[width=0.48\textwidth]{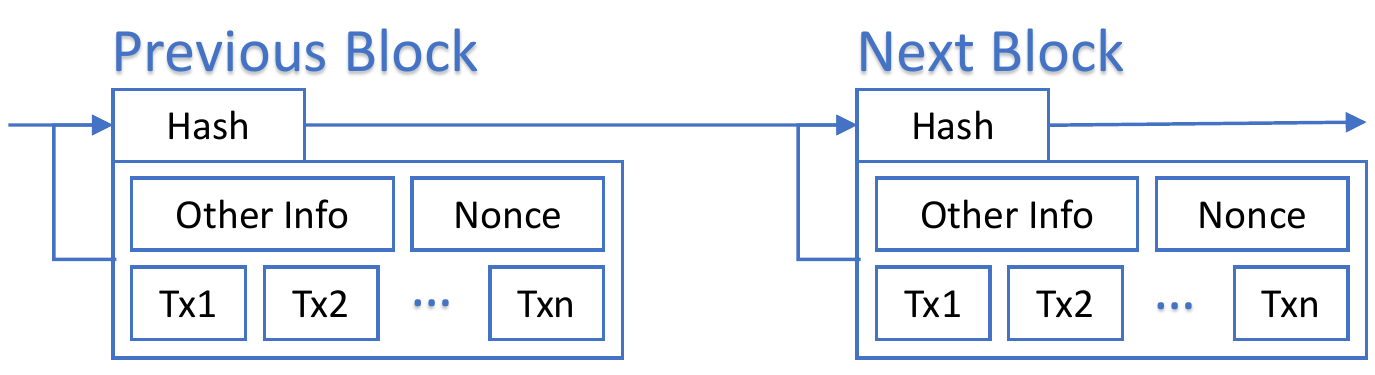}
{The concept of Blockchain.\label{fig:blockchain}}

In a public Blockchain, anyone can participate in the network to create and validate transactions and blocks. Contrarily, only authorized nodes can join a private Blockchain to read, create, or validate transactions and blocks. A consortium Blockchain has mixed features of both private and public Blockchain implementations, where only permissioned users can perform Blockchain transaction with different levels of restrictions. Bitcoin and Ethereum are examples of public Blockchain implementations, and Hyperledger Fabric \cite{Cachin2016ArchitectureOT} is an example of a consortium Blockchain. We can also have a private Blockchain using a private version of Hyperledger Fabric or Ethereum \cite{privateBlockchains2017}. Zheng \textit{et al.} \cite{zheng2018Blockchain} gave a taxonomy on different types and implementations of Blockchain, and the different consensus algorithms used in them. Beside PoW and Proof-of-Stake (PoS) consensus algorithms, Delegated-PoS (DPoS) \cite{larimer2014delegated} and Practical Byzantine Fault Tolerant (PBFT) \cite{castro1999practical} have been also considered in different implementations. However, PoW is still the most secure consensus algorithm despite its huge computational and energy consumption.

The peer-to-peer (P2P) network of Blockchain should have decentralized management of the data that is synchronized among all peers in the network. To keep the data synchronized efficiently, two message transfer protocols are usually adopted between the nodes, i.e. Gossip \cite{Gossip2002} and Kademlia \cite{Kademlia2002}. Bitcoin uses Gossip, which spreads information by communicating only with neighbors, mimicking the spread of epidemic diseases. Ethereum communication protocol on the other hand is inspired by Kademlia, which maintains a distributed hash table that specifies the communicating neighbors for each node. The peers in the network are usually of three types, i.e. core, full, and light nodes. All peers participate in validating and broadcasting transactions and blocks. Core nodes are responsible for network routing, whereas full nodes are responsible for storing the whole Blockchain. Light nodes are only responsible for maintaining users' accounts in resource-constrained devices.

Asymmetric cryptography and zero-knowledge proof \cite{ZeroKnowledge2013} can be used to secure users' data with the help of Blockchain \cite{DataTrans}. Hence, Blockchain can be used as a secure distributed-database system, e.g. medical record system. The patients can ensure data integrity and privacy by giving data access only to specific medical firms. Records from different hospitals and clinics can be obtained in a secure manner without vulnerable central authorities. Peng \textit{et al.} \cite{VQL2019} implementation prevents falsified data retrieval to ensure authenticity, integrity, and efficiency of Blockchain data queries. With the help of smart contracts, automation in such systems mitigate the need for human centric auditing and revisioning. 

Smart contracts \cite{buterin2014next} were proposed in the 1994 by Nick Szabo \cite{Szabo1994Smart} and rediscovered in the context of Blockchain with Ethereum \cite{wood2014Ethereum}. They enforce rules and conditions inside transactions to lower the cost induced by third central parties, such as law firms. These contracts are automated and permanently stored in Blockchain immutable ledger. Running code inside Blockchain using smart contracts adds decentralization to applications in trustless environments. Smart contracts adds automation to network management, security services, and IoT applications. However, Blockchain still suffer many challenges and problems, like scalability \cite{Scaling} and privacy leakage \cite{Deanonymisation2014}. Electricity consumption of PoW and the capitalism problem in PoS algorithm \cite{king2012ppcoin} are some consensus-related problems in Blockchain. Moreover, the financial use of Blockchain still needs much work from legal and law enforcement perspectives \cite{PittsburghTaxReview}.

\subsection{Blockchain Applications}
Zheng \textit{et al.} \cite{zheng2018Blockchain} classified Blockchain applications to finance systems, reputation systems, public and social services, and security and privacy applications. In this survey however, we focus on security \cite{Blockchain2018Intrusion}, AI \cite{Blockchain2019AI}, IoT \cite{MAKHDOOM2019251}, and healthcare \cite{Pirtle2018} applications. We show how Blockchain has led to the development and enhancement of many applications in these domains. We give below brief descriptions of few Blockchain applications that demonstrate the power of Blockchain in enhancing and simplifying humans lives. Such applications and prototypes pave the way for the smart environments of the future we are looking for:

\textit{MedRec} \cite{ekblaw2016case}: Is a distributed Electronic Health Records (EHR) and medical research data. This decentralized medical claim system handles EHRs using Blockchain. It allows patients to easily and securely share their medical records across different health insurance companies, medical institutions, clinics, and pharmacies. It guarantees authentication, confidentiality, accountability for sharing patients' data. Medical stakeholders can be incentivized to play the role of block miners in this system.

\textit{BitAV} \cite{DBLP:journals/corr/Noyes16}: Is a Blockchain-based fast anti-malware scanning application that secures entire networks. It can provide security services in a decentralized manner to computational-limited environments, like IoT networks, given enough RAM and storage. It is 1,400\% faster than conventional antivirus software, and 500\% less in terms of average update propagation flow. It achieves this performance using the P2P network maintenance mechanism inspired by Blockchain consensus.

\textit{OriginChain} \cite{Adaptable2017}: Is an adaptable Blockchain-based traceability system. It is decentralized, transparent, and tamper-proof; it traces the origin of products across complex supply chains. Private data, customer/product information, product certificates and photos are kept off-chain to increase the performance and save space. However, the hashes of that data are kept on-chain to ensure immutability.

\textit{E-Voting}: Are decentralized electronic voting systems that are usually required to scale well to large scale voting \cite{Amr2020Voting}. Yang \textit{et al.} \cite{voting2018Yang} used an Ethereum smart contract to give a prototype of a voting system that provides confidentiality using homomorphic encryption. The eligibility of the voters, and the integrity and validity of their votes can also be verified. Similarly, Khoury \textit{et al.} \cite{Voting2018Khoury} created transparent, consistent, and deterministic Ethereum smart contracts, which can be modified by voting organizers. Voters should pre-register with mobile phone numbers and can only vote once in that voting platform. Hjálmarsson \textit{et al.} \cite{Voting2018Hamdaqa} used a smart contract in a private version of Ethereum to guarantee transparency and privacy. They used Blockchain-as-a-Service to host nationwide elections, but they still need additional measures to support countries with huge population.

\textit{Reputation Systems}: Are decentralized systems for rewards and educational records. Sharples and Domingue \cite{Sharples2016Educational} democratise educational reputations beyond the academic community using a decentralized Blockchain-based framework. It creates a permanent distributed record of intellectual effort and associated reputational reward. It can be also used for crowd-sourced timestamped patenting, i.e. proof of academic, art, and scientific work.

\section{Blockchain \& Internet of Things (IoT)}
\label{sec:IoT}
IoT devices in smart environments should be digitally connected in order to share their data and automate their tasks. These devices are usually made up of sensors and actuators that connect through the internet. Blockchain can be used to increase IoT automation and solve a number of its limitations, including security, privacy, and scalability. That makes Blockchain one of the enabling technologies for IoT networks in smart environments. Using Blockchain for decentralized monetary transaction and digital asset trading is also an enabler for IoT devices in smart environments. Blockchain was also used as a distributed access management system for the ever-growing IoT networks to mitigate the overhead of centralized architectures \cite{BCmeetsIoT}.

IoT is one of the biggest trends in today's innovations \cite{Whitmore2015}. It enables physical devices to communicate through the internet to send/receive data, and can perform actions. IoT has already emerged in humans life in different domains, such as smart home devices \cite{RISTESKASTOJKOSKA20171454}, smart cities \cite{Zhihong2011IoT}, and smart transportation \cite{xie2001electronic}. It has been proven to be well suited for E-business models, specially with the help of Blockchain and smart contracts \cite{Zhang2015IoT, Zhang2017}. As an example, Zhang and Wen \cite{Zhang2015IoT} proposed an E-business architecture to build systematic, highly efficient, flexible, reasonable, and low cost business-oriented IoT ecosystems. In addition, Zheng \textit{et al.} \cite{zheng2018Blockchain} discussed IoT-Blockchain integration and its associated challenges. They identified scalability, ever-growing storage, privacy leakage, and selfish mining as critical problems. They pointed out that big-data analytics and AI can enhance Blockchain-IoT integration and their applications.

Blockchain technology is a good solution to mitigate the problems of traditional central communication and management systems for large scale IoT devices. On one hand, resource limitations of IoT devices, and scalability issues in Blockchain create big problems for Blockchain-IoT integration. On the other hand, the continuous research effort has led to innovative solutions to these problems, such as IoTA \cite{IoTA}. IoTA is not an abbreviation; rather it comes from IoT and the word "iota", which means an extremely small amount. The name reflects its purpose to connect IoT devices through micro/zero value transactions. Shabandri and Maheshwari \cite{IoTAPaper} developed this architecture as a protocol to provide trust in IoT networks. It eliminates transaction fees and the concept of mining to solve both of those problems. The main component of the IoTA is what they called the Tangle, which is a guided acyclic graph (DAG) for transaction storage. Shabandri and Maheshwari \cite{IoTAPaper} demonstrated the performance of IoTA by implementing two IoT applications, namely a smart utility meter system and a smart car transaction system.

ADEPT (Autonomous Decentralized Peer-to-Peer Telemetry) \cite{panikkar7adept} is another example of a Blockchain-based database-like framework for decentralized IoT networks. It is a proof-of-concept that was produced by a collaboration between IBM and Samsung. ADEPT provides a secure and a low-cost interaction mechanism for IoT devices, where devices have the ability to make orders, pay for them, and confirm their shipment autonomously. The underlying technologies behind ADEPT are Ethereum smart contracts, BitTorrent file sharing and TeleHash peer-to-peer messaging. They used a mix of PoW and PoS consensus algorithms to provide secure decentralization for transaction approval.

The huge number of IoT devices creates a burden on the network and raises problems such as data security, privacy, and integrity. Two of the most challenging issues in IoT security are heterogeneity and scalability of IoT devices that are distributed over the network. Blockchain can solve the security, privacy, and data integrity issues in a decentralized manner. Blockchain is also able to create traceable IoT networks, where transaction data are recorded and verified without intermediary management and control \cite{SurveyIoT2020}. Distributed Blockchain-based management of IoT devices can be also performed at the edge of the network to avoid using distant resources \cite{bahga2016blockchain, Huh2017Managing, Sharma2017DistBlockNet}. Blockchain can also provide a distributed digital payment system for IoT devices to reduce the cost induced by third parties.

To protect IoT data in a Blockchain, a hybrid combination of private and public Blockchains is needed \cite{ateniese2019hybrid}. Data privacy is maintained by private management nodes, whereas the consensus algorithm is maintained by public nodes. In addition, traditional Blockchain consensus algorithms are not suitable for IoT-Blockchain solutions because of their computational and time requirements. To solve this problem, Samaniego and Deters \cite{Samaniego2016} proposed a Blockchain-as-a-service platform for IoT applications. They introduced structural improvements for Blockchain to fit IoT networks by improving consensus algorithms.

Atlam \textit{et al.} \cite{soton421529} listed several benefits for using Blockchain for IoT, like publicity, decentralization, resiliency, security, speed, cost saving, immutability, and anonymity. However, they also highlighted the challenges, such as scalability, processing power, time delay, storage requirements, lack of skills, legal and compliance, and naming and discovery. Similarly, Ramachandran and Krishnamachari \cite{ramach2018blockchain} proposed Blockchain-based monetary exchange for data and compute in IoT networks. IoT transactions can also be recorded on Blockchain for future accounting and auditing. However, they were also concerned by Blockchain challenges, like latency, bandwidth consumption, transaction fees, transaction volumes, partition tolerance, and physical attacks on IoT devices.

Dorri \textit{et al.} \cite{dorri2016blockchain} investigated the delay, expensive computations, and bandwidth overhead problems of Blockchain to better fit IoT applications. They proposed a secure, private, and lightweight hierarchical architecture for Blockchain-IoT applications. The hierarchical architecture is made up of three layers, namely local network (smart home), overlay network, and cloud storage (see Fig. \ref{fig:DorriArch}). Likewise, Reyna \textit{et al.} \cite{REYNA2018173} argued the benefits of Blockchain-IoT integration to securely push code into IoT devices to speed up the deployment of new IoT ecosystems. They proposed Blockchain-based direct firmware update without the need to trust third-parties. Since IoT devices are manufactured by different vendors, they may not agree on sharing a common Blockchain. Hence, IoT devices should be able to send transactions across different Blockchain implementations with different consensus protocols.

\Figure[t!](topskip=0pt, botskip=0pt, midskip=0pt)[width=0.48\textwidth]{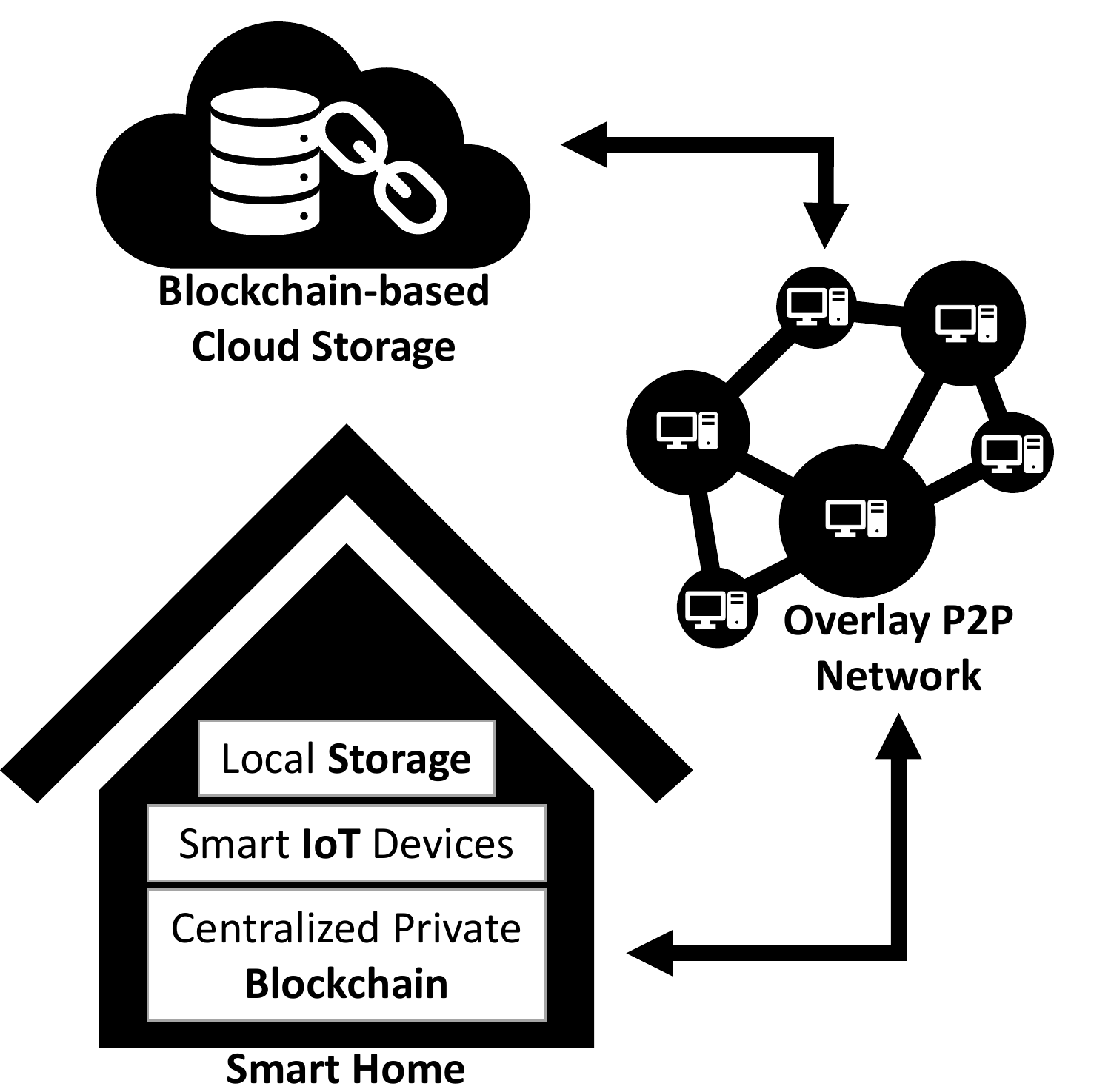}
{Hierarchical Blockchain-based IoT architecture for smart homes.\label{fig:DorriArch}}

Dai \textit{et al.} \cite{BCIoTSurvey} named the integration of Blockchain and IoT as Blockchain-of-Things (BCoT). They stated that a successful integration requires interoperability, traceablitiy \cite{Adaptable2017}, reliability, and autonomicity \cite{Zhang2015IoT}. They also stated that decentralization, heterogeneity, poor interoperability, privacy and security vulnerabilities \cite{Christidis2016BusinessModel} are critical issues for such integration. Beside those problems, the lack of publicly available IoT datasets for the research community is another problem to be addressed. Thus, Banerjee \textit{et al.} \cite{BANERJEE2018149} worked on standards for securely developing and sharing IoT datasets. They proposed two conceptual solutions to ensure IoT data integrity and privacy using Blockchain.

Reyna \textit{et al.} \cite{REYNA2018173} proposed using three different communication approaches for the communication between IoT devices with the help of Blockchains (see Fig. \ref{fig:ReynaArch}). The first approach is IoT-IoT interactions, where IoT interactions take place off-chain. This approach is the fastest among other approaches since only a part of IoT data is stored on-chain. The second approach is IoT-Blockchain, where all the interactions take place through Blockchain. With this approach, all IoT data are stored on-chain to ensure traceable interactions. The third approach is hybrid, where only part of the interactions/data goes through Blockchain, while the rest is done directly between IoT devices. The hybrid approach is better in terms of performance and security; however, it requires careful orchestration for those interactions. 

\Figure[t!](topskip=0pt, botskip=0pt, midskip=0pt)[width=0.48\textwidth]{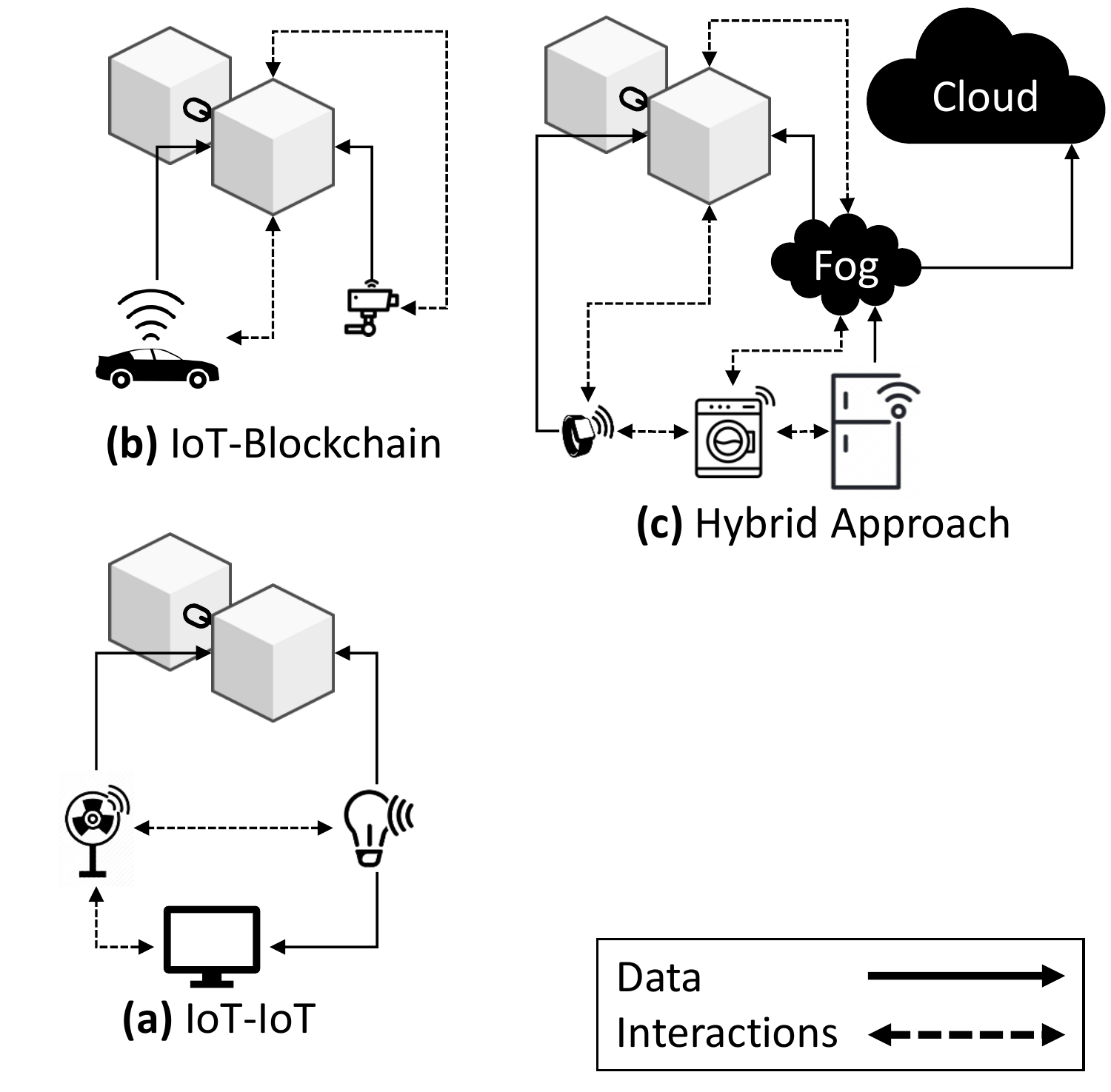}
{Three different Blockchain-IoT interaction architectures proposed by Reyna \textit{et al.} \cite{REYNA2018173}.\label{fig:ReynaArch}}

The impact of IoT on industry and enterprise systems asked for standardizing this technology to speed up its development and spread in this domain \cite{IoT2014Xu}. Christidis and Devetsikiotis \cite{Christidis2016BusinessModel} showed that Blockchain-IoT integration will cause transformations across industries, and open the door for new business models and distributed applications. It will also facilitate service and resource sharing, and automate time-consuming workflows. Blockchain smart contracts are the main source of such automation for complex multistep processes. They can reduce cost and time for future business models and applications in smart environments. 

\subsection{Blockchain-IoT Applications}
A number of key applications were developed for smart environments with the help of Blockchain-IoT integration. Prototypes of these applications are necessary to study them and solve their limitations and issues. Here, we list few applications, case studies, and prototypes for smart environments that are brought to live using Blockchain-IoT integration:

\textit{Smart Homes}: Dorri \textit{et al.} \cite{dorri2016blockchain} proposed a smart-home case study that uses Blockchain to add security and privacy features to various IoT applications. They presented a lightweight secure system for IoT-based smart homes to minimize the overhead of consensus algorithms. Their hierarchical architecture consists of smart homes, an overlay network, and cloud storage. Other works later analyzed this architecture and the role of smart homes as miners in a private Blockchain \cite{Dorri2017SmartHome, OptBC4IoT}. They used simulation results to show the insignificance of Blockchain overhead, including power consumption, compared to achieving confidentiality, integrity, availability, security, and privacy.

\textit{Energy Trading}: Sikorski \textit{et al.} \cite{SIKORSKI2017234} presented a proof-of-concept and a detailed implementation for energy trading using realistic data. It is a machine-to-machine electricity market for the chemical industry. They used a Blockchain smart contract for automatic confirmation of trading and payment commitments. They implemented a scenario of two electricity producers and one consumer that automatically trades energy using IoT. Producers publish energy trading offers for a given price, while consumers can read, analyse, and accept, or refuse those offers. Consumers can pick and accept the offer with the minimum cost using a smart contract execution as an atomic exchange of assets, i.e. currency for energy. Each transaction is saved on the immutable ledger for future proofing.

\textit{Smart Things}: Panarello \textit{et al.} \cite{Integeration2018Penarello} did a systematic survey of Blockchain-IoT integration. They covered various smart-application domains, such as smart homes, smart properties, and smart cities. They also covered smart energy-trading, smart manufacturing, smart data-marketplaces, and other generic smart applications. They classified existing work based on the development levels, consensus algorithms, and technical challenges. They also identified the challenges that include confidentiality, authentication, integrity, availability, and nonrepudiation.

Conoscenti \textit{et al.} \cite{systematic2016Conoscenti} gave a systematic literature review of Blockchain applications for IoT. They discussed 18 use cases of Blockchain, 4 of which are specifically designed for IoT. The rest of the use cases are applications for decentralized private data management systems that are inline with IoT applications. The four IoT-related Blockchain applications they discussed are:

\begin{enumerate}
    \item \textit{E-business models for IoT solutions}: Zhang and Wen \cite{Zhang2015IoT} designed a methodology for transactions and payments between smart IoT devices using Blockchain smart contracts.
    
    \item \textit{IoT Data-Market}: Wörner and von Bomhard \cite{sensor2014money} proposed a prototype for a system where sensors can sell data directly to a data-market in exchange of Bitcoins.
    
    \item \textit{Public-Key Infrastructure}: Axon and Goldsmith \cite{axon2016a} adapted what is called Certcoin \cite{CertCoin} to a privacy-aware Blockchain-based public-key infrastructure to avoid web certificate authorities, provide certificate transparency, and mitigate single points of failures.
    
    \item \textit{Enigma}: An autonomous Blockchain-based decentralized computation platform proposed by Shrobe \textit{et al.} \cite{Enigma2018}. It allows different users to run computations on personal data with guaranteed privacy.
\end{enumerate}

In order to fully benefit from IoT applications and systems, we need to start by mitigating their current limitations and challenges. The power and resource limitations of IoT devices do not make them suitable to process heavy computations and store large data. Cloud Computing helped providing theoretically unlimited storage and computational resources for IoT devices. In addition, Fog and Edge Computing bring those resources closer to IoT devices to decrease network delay and bandwidth consumption. Cloud, Fog, and Edge resources allow for performing heavy analysis on IoT data and maintain real-time performance for time-sensitive IoT applications. These resources will extend the capabilities of Blockchain-IoT integration and mitigate many of its limitations and issues.

\section{Blockchain \& Cloud, Fog, and Edge Computing}
\label{sec:Cloud}

The huge amount of IoT data generated in smart environments needs to be processed in large data centers that have enough computing and storage capacities. That is why Cloud Computing was proposed as the first solution for big data analysis and storage for IoT-based applications in smart environments \cite{Cloud_Big_2021}. However, Fog Computing paradigm was evolved to support the Cloud in order to mitigate latency intolerance of real-time IoT applications in smart environments \cite{Cloud_2Fog_2021}, such as autonomous vehicles. Similarly, Edge Computing utilizes available computational resources in smart environments, such as the resources in smart vehicles or smart phones, in order to reduce the latency even further \cite{Cloud_Fog_2021}. Blockchain was again able to support these technologies by securing and protecting the privacy of this big data in smart environments \cite{Cloud_BC_2021}. 

With the advent of the Cloud Computing technology, it is possible to perform very expensive computational tasks and store tremendous amount of data. Payment is only for the cost of usage, which is better than purchasing expensive resources for time-framed tasks. The Cloud technology removes the overhead of maintenance and resource management, which is usually related to owning resources by small to medium sized companies. Furthermore, the Cloud enables resource and power-limited devices, such as smartphones and IoT devices, to perform heavy computations and store huge amounts of data. Those devices need to only use a lightweight remote interface with the cloud as a solution to mitigate their power and resource limitations.

Cloud Computing is the outcome of integrating parallel, distributed, and grid computing \cite{Zhang2010Cloud}. Although it has been proposed in the 60s, the technology has started to be widely used commercially in 2006 by Amazon \cite{Jadeja2012Cloud}. Services are usually provided as packages, namely Software-as-a-Service (SaaS), Platform-as-a-Service (PaaS), and Infrastructure-as-a-Service (IaaS). Zhou \textit{et al.} \cite{Zhou2010Cloud} added the possibility of having Data-as-a-Service (DaaS), Identity and Policy Management as-a-Service (IPMaaS), Network-as-a-Service (NaaS). X-as-a-Service (XaaS) is a term called for the countless number of services that can be provided by cloud computing \cite{Ma2012Cloud}. XaaS allows for more service packages, which enable the creation of various applications and systems based on the services provided.

Jadeja and Modi \cite{Jadeja2012Cloud} categorized the deployment of the cloud infrastructures into public, private, hybrid, and community clouds. They listed easy management, cost reduction, uninterrupted services, disaster management, and green computing as the main advantages of Cloud Computing. A tremendous number of systems and applications were built based on Cloud services since 2010 \cite{Zhou2010Cloud}. Ma and Zhang \cite{Ma2012Cloud} studied a provider for cloud services called Google App Engine (GAE). They explored three of its services, i.e. Google File System (GFS), MapReduce, and Bigtable. They showed how these services opened the doors for Big Data Analysis in IoT environments.

Before discussing how Cloud integrates into Blockchain-based IoT solutions, we first show how Blockchain helped the Cloud technology itself. Blockchain has been used for Cloud Exchange \cite{BlockchainSurvey2020}, which allows for provisioning and management of multiple Cloud providers. It can lower the price and can provide flexible options for Cloud users. Xie \textit{et al.} \cite{BlockchainSurvey2020} proposed using Blockchain to decentralize Cloud exchange services. It mitigates malicious attacks and the cheating behaviors of third-party auctioneers. Furthermore, Blockchain form new models for security-aware Cloud schedulers, like the lightweight Proof–of–Schedule (PoSch) consensus algorithm \cite{WILCZYNSKI2020Modelling}. In addition, integrating the Cloud with Blockchain-IoT solutions enables for seamless authentication, data privacy, security, easy deployment, robustness against attacks, and self-maintenance \cite{Review2018}. 

Because of limited power, storage, and computational resources, IoT devices heavily depend on Cloud resources. However, relying on the cloud can create unacceptable delays, specially when a feedback is required. Fog and Edge technologies can be used to reduce these delays; in addition, they provide better privacy by processing IoT data in proximity to IoT devices. Therefore, they are more convenient than the Cloud, specially for Blockchain-IoT applications. Samaniego and Deters \cite{Samaniego2016} evaluated both edge-based and cloud-based Blockchain implementations for IoT networks. They did show, via simulations, that edge-based Blockchain implementations outperform cloud-based implementations.

Transferring massive amounts of data, that is produced by IoT devices, to the Cloud consumes considerable amount of network bandwidth. Flooding the core network with massive traffic, like in streaming IoT applications, will create network bottlenecks and single points of failures. There are also problems with privacy exposure, and context and geographical location unawareness. Edge Computing solves these problems by forming a distributed and collaborative computing resources. It reduces power consumption, provides real-time service, and improves scalability for many industrial IoT solutions \cite{Chen2018Edge}. In addition, Unmanned Aerial Vehicles (UAVs) can act as limited-resource edge servers in environments with limited or no infrastructures \cite{UAV2018}. Blockchain can be used to provide mutual-confidence between UAVs of different providers \cite{UAV2017Kapitonov}, and to preserve privacy and security for their data \cite{kumar2018unmanned}.

Fog Computing has been also proposed to pre-process and trim IoT data before sending it to the cloud for computationally expensive analysis \cite{FogComputing2014}. Fog servers are usually deployed in smart gateways, which are equipped with decent computational resources. They eliminate unnecessary communication to the cloud to save the network bandwidth and reduce the load in its data centers. Fog Computing should not be mixed with Edge Computing, as Edge Computing brings the computations very close to the end devices. Edge servers are usually deployed in Radio Access Networks (RANs) or mobile Base Stations (BSs). On the other hand, Fog Computing provides distributed mini-cloud resources between the end devices and the cloud (see Fig. \ref{fig:FogEdge}). Both technologies reduce the network delay, and provide better Quality-of-Service and Experience (QoS \& QoE). These technologies are essential parts in real-time/streaming IoT applications, which might also require location/context-aware information processing.

\Figure[t!](topskip=0pt, botskip=0pt, midskip=0pt)[width=0.48\textwidth]{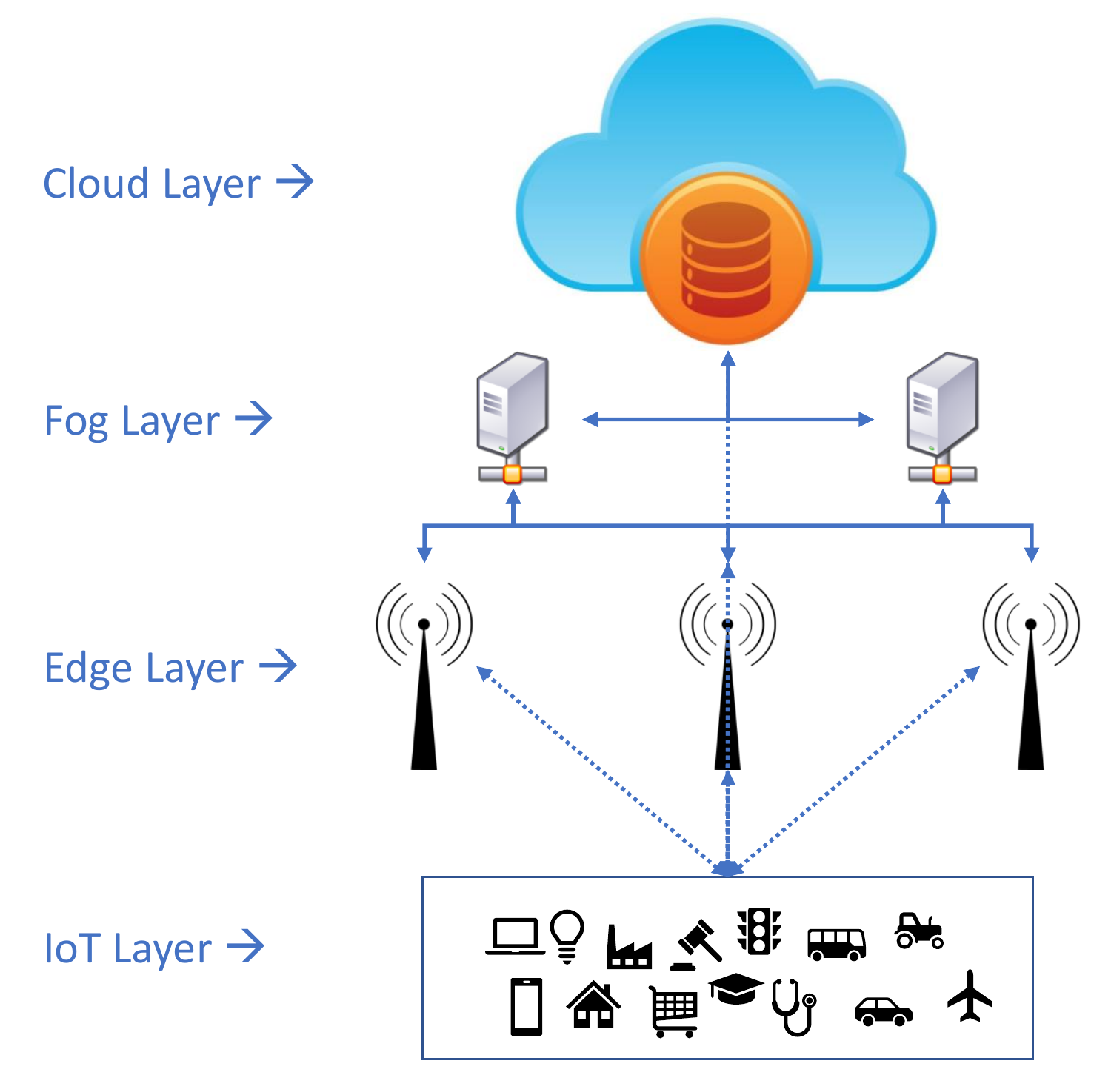}
{Cloud, Fog, and Edge Computing for IoT networks.\label{fig:FogEdge}}

Mobile Edge Computing, also called Multiaccess Edge Computing (MEC), is a specific type of Edge Computing that leverages mobile Base Stations. It complements cloud computing by offloading computations closer to mobile and IoT devices. MEC supports ultralow latency and delay-sensitive IoT applications in 5G networks \cite{Abbas2018MEC}. Xiong \textit{et al.} \cite{Xiong2018MobileBlockchain} proposed a prototype for MEC-enabled Blockchain for mobile IoT applications. However, the limited MEC resources makes it critical to optimally offload heavy computations to different Edge, Fog, or Cloud resources. Liu \textit{et al.} \cite{Liu2018Offloading}, for example, optimized the joint computation offloading and content caching problems to tackle the intensive computations in PoW consensus algorithm.

Xiong \textit{et al.} \cite{Xiong2019CFC} studied the relationship between cloud or fog providers and PoW-based Blockchain miners with limited computation resources. They chose to offload the computational intensive part of PoW to the cloud and/or fog nodes. The computing nodes offer services to the miners for a given price using a game theoretic approach. The miners can then decide on the amount of service to purchase from the computing nodes. Tuli \textit{et al.} \cite{TULI201922} also used Blockchain to provide authentication and encryption services to secure IoT sensitive data and operations. They proposed FogBus, a lightweight end-to-end platform-independent framework for IoT applications. It enables easy deployment, scalability, and cost efficiency by integrating IoT into cloud, fog, and edge computing with the help of Blockchain. 

Blockchain has been used to provide distributed access control for IoT devices \cite{BCIoTAccess}. Almadhoun \textit{et al.} \cite{Almadhoun2018Authentication} proposed a user authentication system for IoT devices using fog computing. In their proposed system, fog nodes utilize an Ethereum smart contract to authenticate the users and manage access permissions. The proximity of fog nodes to IoT devices provides real-time services for the users. To mitigate malicious attacks on fog nodes, Wu and Ansari \cite{Wu2020Cooperative} proposed partitioning the fog nodes into different clusters. The nodes in each cluster have their own access control list, which is protected and managed by Blockchain. They showed, using simulations, the effectiveness of their approach in reducing the computational and storage requirements of Blockchain. They included a heuristic algorithm to reduce the time needed to solve the consensus puzzle by having all fog nodes perform the computations cooperatively.

Cloud, Fog, and Edge computing technologies provided IoT devices with more capabilities, which increased their adoption in new applications and domains. The continuous expansion of IoT networks and their dynamic nature ask for intelligent and dynamic management of such networks. Adaptive control of the network does not only save in terms of hardware cost, but also dynamically optimizes the operations in the network. Network optimization can be done using Software-Defined Networking, which dynamically changes data flow in the network based on its state. SDN can incorporate Artificial Intelligence algorithms to optimally choose between using Cloud, Fog, or Edge resources, or even a combination of them, to process IoT data.

\section{Blockchain \& Software-Defined Networking (SDN)}
\label{sec:SDN}

SDN Technology demonstrated its importance in managing routing decisions in smart environment IoT networks since they are usually vulnerable to node/link failures \cite{SDN_Routing_2021}. SDN has been used to mitigate latency issues, like congestion and transmission delays, in time-sensitive smart industrial IoT environments \cite{SDN_Time_2021}. SDN also balances the load between Fog nodes, which can be vehicles in IoV environments, and the Cloud to allow time-sensitive tasks meet their deadlines \cite{SDN_Cars_2021}. When a Fog is overloaded, SDN dynamically makes offloading decisions to select the best offloading Fog node based on computational and network resource information \cite{SDN_Fog_2021}. Blockchain augments SDN benefits for IoT networks by providing security, privacy, flexibility, scalability, and confidentiality to increase energy utilization and throughput while reducing end-to-end delay \cite{SDN_BC_2021}.

SDN enables network management and protocols to be adaptable and programmable. It was proposed in 2006, in the OpenFlow whitepaper, to test experimental protocols in university campus networks \cite{openflow2008}. SDN fits to the rapid changes and demands in network applications, and eliminates the need for pre-programmed, vendor-specific, and expensive network devices. SDN achieves this flexibility by separating the control and data planes in the network. The control plane makes decisions on the traffic flow in the network, whereas the data plane is responsible for forwarding that traffic. Indeed, the control plane is the network brain, and is usually a centralized software entity called SDN controller.

The SDN architecture uses the concept of Application Programming Interface (API) in a three-layer structure. Fig. \ref{fig:SDN} shows two common interfaces between these layers, i.e. Northbound and Southbound APIs. Two additional interfaces are sometimes considered to allow the communication between multiple controllers in the control layer, i.e. Eastbound and Westbound APIs \cite{SDN2017Survey}. A master controller is usually needed to coordinate the decisions of multiple controllers. A controller is the interface between network elements and applications like firewalls and load balancing applications \cite{Badotra2020}. It provides agility to network infrastructures, like routers and switches, by dynamically optimizing the network resources. Like other emerging technologies, SDN introduces new security challenges to network infrastructures. At the same time Blockchain can mitigate those challenges by providing confidentiality, integrity, and availability to the network devices \cite{Alharbi2020SDNBlockchain}.

\Figure[t!](topskip=0pt, botskip=0pt, midskip=0pt)[width=0.48\textwidth]{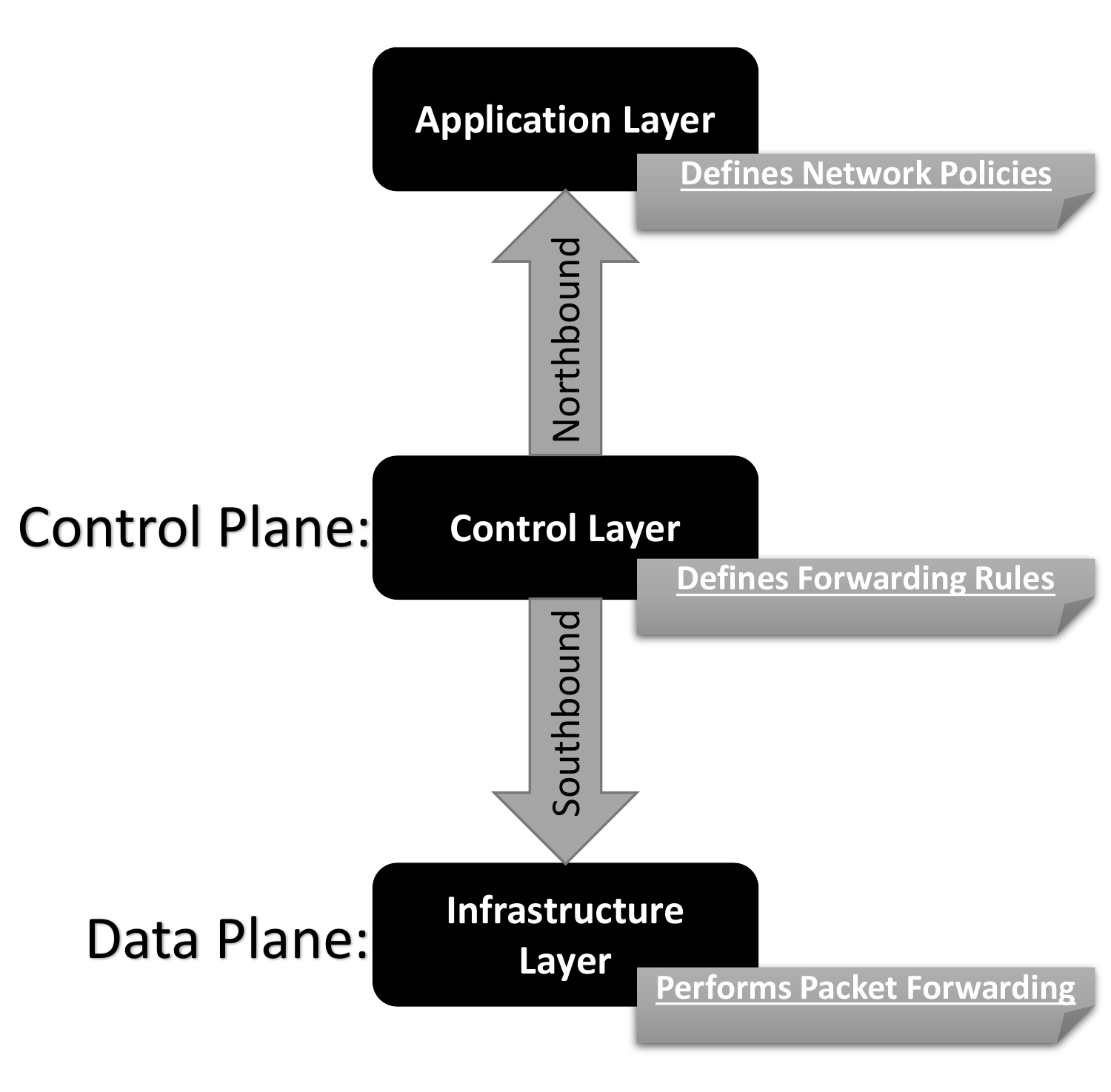}
{The three-layers SDN architecture.\label{fig:SDN}}

Denial of Service Attacks (DoS) and Distributed Denial of Service Attacks (DDoS) target centralized network architectures. These attacks stop network services from serving legitimate users, devices, and applications. Such attacks can target SDN controllers to malfunction and paralyze the whole network. Blockchain can mitigate these attacks and help avoiding single-points-of-failures in centralized network architectures. Blockchain can provide decentralized trust in physically distributed, logically centralized, SDN controller architectures \cite{Alharbi2020SDNBlockchain}. Using SDN and Blockchain, network administrators can easily program and configure network components using smart contracts. Those components can securely perform their software updates by accessing policies and configurations from Blockchain-based SDN controllers.

Blockchain integration with SDN did not attract the required level of attention yet by the research community \cite{Alharbi2020SDNBlockchain}. Alharbi \cite{Alharbi2020SDNBlockchain} explained the role of Blockchain in providing and improving security features in SDN architectures. The dynamism, adaptability, and remote configuration of SDN networks provide a lot of support for IoT networks. Bera \textit{et al.} \cite{SDN2017Survey} showed how these features can provide efficient, scalable, seamless, and cost-effective management of IoT devices. SDN also realizes the real-time demands of IoT applications by its ability to optimize traffic flow and load balancing. Such optimization improves the bandwidth utilization in the network and mitigates bottlenecks.

Jararweh \textit{et al.} \cite{Jararweh2015} proposed a comprehensive SDN-based IoT framework to simplify IoT management and mitigate several problems in traditional IoT architectures. They integrated software-defined networks, storage, and security into a single software-defined control model for IoT devices. The software-defined storage manages big IoT data by separating the data control layer, which controls storage resources, from the underlying infrastructure of storage assets. Finally, the software-defined security separates the data forwarding plane from the security control plane. They included a proof-of-concept to show the performance of their framework in handling huge amounts of IoT data.

The global network view in SDN controllers addresses the heterogeneity, scalability, optimal routing, and bottleneck issues in IoT networks. Kalkan and Zeadally \cite{SDN2018Securing} discussed the benefits and drawbacks of SDNs in IoT networks and focused on single-point-of-failure issues in traditional centralized SDN controllers. They separated the roles of single controllers to multiple hosts using a distribution-of-risks scheme. The bandwidth utilization was improved by distributing the communication traffic among three different controllers, i.e. Intrusion, Key, and Crypto Controllers. The intrusion controller mitigates possible intrusions besides managing and securing the routes. The key controller controls symmetric and asymmetric key distribution in the whole ecosystem. The crypto controller provides cryptographic services for authentication, integrity, confidentiality, privacy, and identity management.

LI \textit{et al.} \cite{Wenjuan2020BlockchainSDN} focused on the security challenges and solutions for Blockchain-based SDN systems. They focused on DoS and DDoS attacks on centralized SDN controllers, and insider attacks on distributed SDN controllers. They highlighted Blockchain ability to secure distributed SDN controllers and data plane forwarding devices. They also listed scanning, spoofing, hijacking, DoS, and Man-in-the-middle attacks as SDN vulnerabilities. At the end, they listed traffic-flow control, policy enforcement, and DoS defence mechanisms as possible solutions for those vulnerabilities. Blockchain was also used to ensure the security and consistency of the statistics in SDN-based IoT networks \cite{Huo2020}.

Medhane \textit{et al.} \cite{Medhane2020EdgeCloudSDN} proposed a security framework for next generation IoT by integrating Blockchain with SDN, edge, and cloud computing technologies. The framework features security attack mitigation, continuous confidentiality, authentication, and robustness. The attacks in IoT networks are detected in the cloud and reduced at the edge nodes. SDN controllers examine and manage the traffic flow to actively mitigate doubtful traffic. Similarly, Sharma \textit{et al.} \cite{Sharma2017IoT} described a distributed cloud architecture at the edge of the network; i.e. Fog nodes, which is secured using Blockchain and SDN. They securely aggregate IoT data using fog nodes before it is sent to the cloud for heavier analysis and/or long-term storage. Their architecture supports large IoT data using low-cost, high-performance, and on-demand secure services. They significantly reduced traffic load and delay compared to traditional IoT architectures. 

Sharma \textit{et al.} \cite{Sharma2017DistBlockNet} proposed DistBlockNet, a decentralized and secure Blockchain-based SDN architecture that updates flow-rule tables in large-scale IoT networks. Blockchain was used to verify flow-rule tables' versions, and securely download them to IoT/forwarding devices. Sharma \textit{et al.} \cite{Sharma2018SDN} also proposed SoftEdgeNet to extend their previous works \cite{Sharma2017DistBlockNet, Sharma2017IoT}. SoftEdgeNet improved their previous designs by pushing the storage and computations to the extreme edge to manage real-time traffic and avoid resource starvation. It is a distributed network management architecture for edge computing networks. It mitigates flooding attacks and provides real-time network analytics using Blockchain, SDN, fog, and edge nodes. SoftEdgeNet has an efficient flow-rule allocation and partitioning algorithm at the edge of the network that minimizes traffic redirection, and creates a sustainable network.

Blockchain has been also used to secure the configuration, management, and migration of Virtual Network Functions (VNFs) \cite{VNF2018}. VNF, also called Network Function Virtualization (NFV), allows devices with adequate resources to perform multiple tasks simultaneously, or at least in real-time. VNF achieves multitasking by separating the control plane from the physical devices \cite{SDN2015NFV}. Alvarenga \textit{et al.} \cite{VNF2018} implemented a prototype that makes VNF configuration immutable, auditable, nonrepudiable, consistent, and anonymous. Their design eliminates single-points-of-failures and provides high availability of the network's configuration information with a delay of about two seconds. The architecture is resilient to Blockchain collusion attacks, and the configuration information cannot be compromised even with a successful collusion attack. Such resiliency is achieved by using a variant of the Byzantine Fault Tolerant (BFT) consensus protocol called Ripple protocol \cite{schwartz2014ripple}.

Blockchain was also used in wireless network virtualization ecosystems to prevent double spending of wireless resources at a given time and location \cite{fusion2019, PoWR}. Wireless network virtualization enables sharing physical wireless infrastructures and radio frequency slices to improve coverage, capacity, and security. Proof-of-Wireless-Resources (PoWR) \cite{PoWR} has been proposed to mitigate double spending of the same wireless resources. Rawat \cite{fusion2019} used SDN to provide dynamic and efficient network configuration, and used Edge Computing to decrease delays by avoiding the use of high-speed backhauls. Such fusion of Blockchain with SDN and edge computing guarantees QoS for end users, and provides trust, transparency, and seamless subleasing of resources in trustless wireless networks. The use of Blockchain makes it practically infeasible to create malicious attempts to sublease other's wireless resources.

The dynamic capabilities of SDN-based networks open the doors for many applications, specially in IoT ecosystems. These capabilities can be further enhanced by adding intelligent decision making into their controllers. Such intelligent decisions are now possible to be inferred using AI and machine learning algorithms. With the advent of Deep Neural Networks (DNN), these algorithms can learn on highly complex environments. Recent DNN-based algorithms achieved accuracies that exceed human abilities in different domains. SDN controllers can deploy DNN-based Reinforcement Learning algorithms to create intelligent agents that dynamically adapt to network changes. These capabilities are a must for IoT networks in future smart environments.

\section{Blockchain \& Artificial Intelligence (AI)}
\label{sec:AI}

AI and Machine Learning algorithms play an essential role in adding automation and intelligence into different smart environment applications, including smart cities applications \cite{AI_Smart_2021}. Besides supporting smart applications, AI also augments various underlying technologies, like optimizing SDN monitoring to minimize network latency \cite{AI_RL_2021}. Furthermore, Blockchain empowers AI decision-making by making it more secure and efficient \cite{AI_BC_2021}. For example, integrating Blockchain and AI provides decentralized authentication for smart cities \cite{AI_Cities_2021}, where user identities are kept secret while attackers are automatically identified. In smart health systems, Blockchain integration with AI helped secure medical data sharing \cite{AI_Hospitals_2021} and protect personal healthcare records \cite{AI_Health_2021}. For smart energy trading, Blockchain-enforced Machine Learning predictive analysis models provide real-time support and monitoring as well as immutable transaction logs for decentralized trading \cite{AI_Energy_2021}. Likewise, Blockchain integration with Machine Learning in smart factories can secure system transactions and deliver smarter quality control schemes \cite{AI_Manufact_2021}.

Akter \textit{et al.} \cite{Akter2020} defined the ABCD of digital business as AI, Blockchain, Cloud, and Data analytic. They considered these emerging technologies as transformation factors for future digital business models. For successful digital business, Garcia \cite{Garcia2020} proposed a complete legal doctrine for smart digital economy. Technologies like Blockchain, AI, IoT, and big data guide governments’ annual budget plans to simplify and maximize the application of taxes for digital businesses. Garcia also showed that Blockchain and cryptocurrencies augments AI and IoT to create smart economy in smart digital world. Ekramifard \textit{et al.} \cite{Ekramifard2020} studied how AI algorithms improve Blockchain designs and operations. They discussed the effect of this integration in the medical field, like the ability to gather, analyse, and make decisions on medical datasets. AI and Blockchain helped in different medical applications, including systems for the ongoing COVID-19 pandemic \cite{nguyen2020, Mashamba_Thompson_2020}. Mashamba-Thompson and Crayton \cite{Mashamba_Thompson_2020} integrated Blockchain and AI to create a low cost self testing and tracking system for COVID-19. Their system is ideal in environments with poor access to laboratory infrastructures. Similarly, Nguyen \textit{et al.} \cite{nguyen2020} discussed how Blockchain and AI were used in the literature to compact the COVID-19 pandemic.

Al-Garadi \textit{et al.} \cite{AI4IoT} discussed the role of machine learning and deep learning in IoT security. They suggested integrating Edge computing and Blockchain into machine learning and deep learning to provide reliable and effective IoT security methods. In their work, they investigated the use of Neural Networks (NN) and other machine learning algorithms (see Fig. \ref{fig:DL}) to detect attacks in IoT networks. Based on IoT and network data, IoT network state is classified into normal (secure), early warning, or attacked state. Fig. \ref{fig:DL} shows a brief taxonomy of different algorithms in the field of AI that can be adopted in IoT ecosystems. Those algorithms can be classified into classical machine learning algorithms and algorithms based on DNN. Each of those classes can be further categorized into supervised, unsupervised, and semi-supervised methods. The supervision in learning algorithms means introducing training labels with training examples that were prepared by professional or computer software. It is sometimes impractical or hard to create labeled datasets; in this case, we can use unsupervised algorithms to categorize and cluster the data into different groups. Semi-supervised algorithms lay between those two classes, where only a small portion of the data is labeled, whereas there are no labels for the rest of the data.

\Figure[t!](topskip=0pt, botskip=0pt, midskip=0pt)[width=0.48\textwidth]{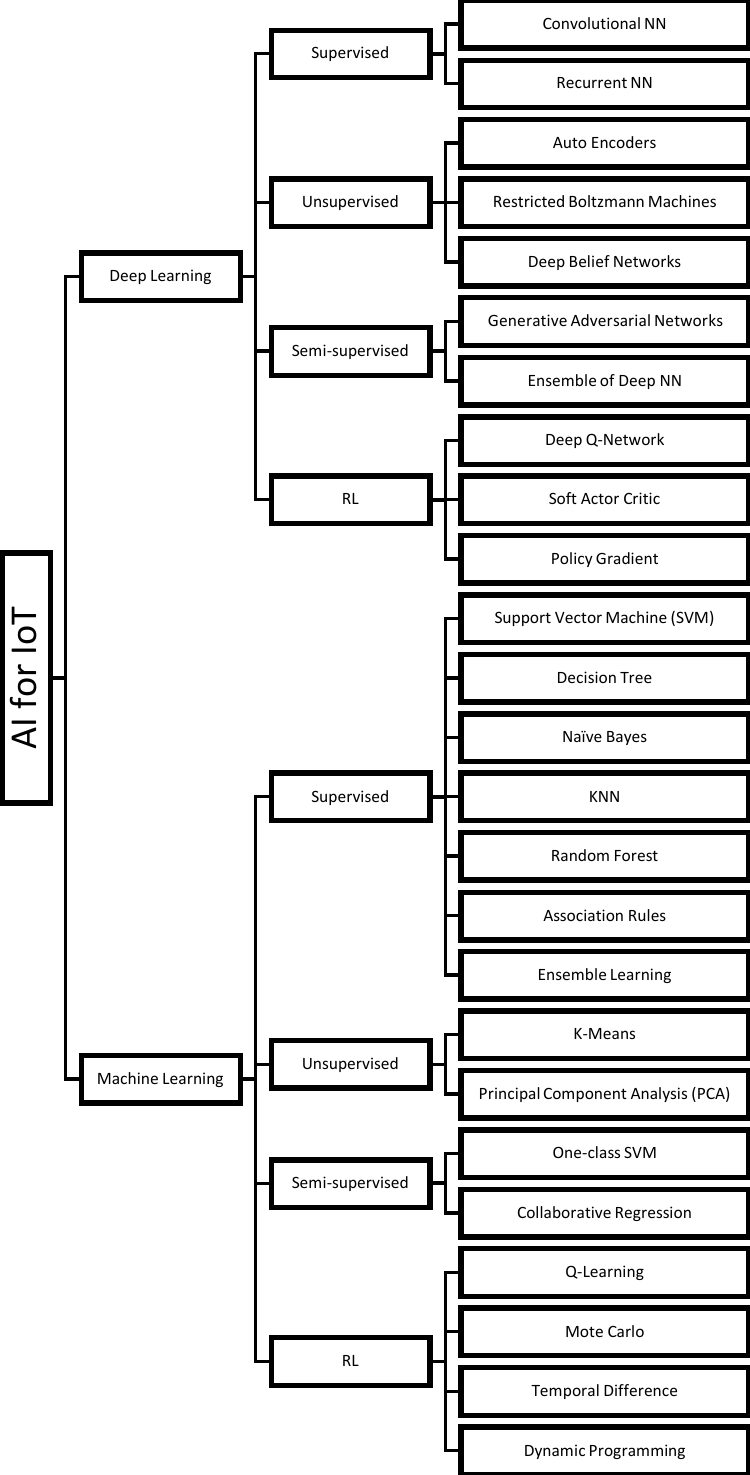}
{A taxonomy of Artificial Intelligence algorithms for IoT ecosystems.\label{fig:DL}}

AI integration is essential to provide smart decision-making capabilities into different technologies in a smart environment ecosystem. The breakthroughs in machine learning and Deep Learning algorithms make them suitable for solving complex problems in rapidly changing environments, such as IoT networks. AI is important to enhance the performance of technologies, like Blockchain, IoT, SDN, Cloud, Fog, and Edge Computing. Self-driving vehicles, smart transportation, automatic delivery robots are some examples of smart environment applications that need AI integration into Blockchain-IoT solutions. AI can also be used to optimize the global energy consumption in a smarter and greener world to decrease the effect of climate change and local air pollution. Kumari \textit{et al.} \cite{KUMARI2020148}, for example, studied the advantages and challenges of integrating Blockchain with AI in Energy Cloud Management (ECM) systems. Using IoT and Smart Grids (SG), this integration allows for sustainable energy management and efficient load prediction in a trustless environment (see Fig. \ref{fig:energy}). They also proposed a decentralized Blockchain-based AI-powered ECM framework for energy management to mitigate security and privacy issues in traditional implementations.

\Figure[t!](topskip=0pt, botskip=0pt, midskip=0pt)[width=0.48\textwidth]{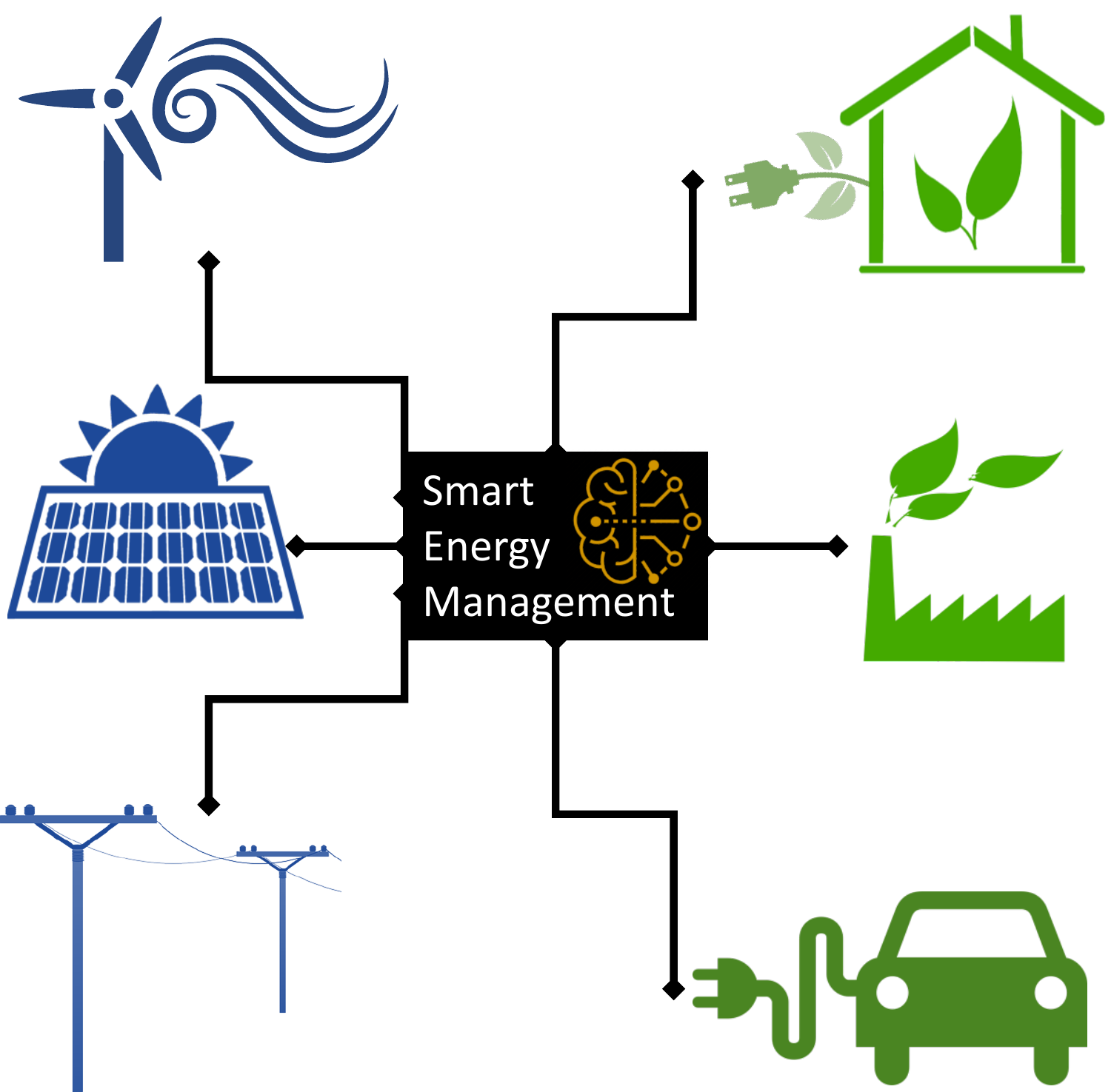}
{Smart energy management for sustainable energy and green smart environments.\label{fig:energy}}

AI can provide best pricing for IoT data to be sold and/or computations to be performed. AI can also empower SDN controllers to choose optimal network routes to forward data traffic. AI-based SDN traffic control can minimize network delay and bandwidth consumption. It is challenging to optimally offload computations between end-devices and the Cloud, Fog, and/or Edge servers jointly considering delay, computations, and power resources. AI algorithms, like Deep Reinforcement Learning (DeepRL), have been recently considered for task offloading and orchestration in edge computing applications \cite{Dai2019AI, BlockchainRL2019}. Dai \textit{et al.} \cite{Dai2019AI} used DeepRL to dynamically orchestrate edge computing and caching resources in complex vehicular networks. The complexity of such networks comes from vehicles mobility and content popularity/localization, e.g. a car accident at a given location. Furthermore, Dai \textit{et al.} \cite{BlockchainRL2019} proposed to integrate AI and Blockchain to provide intelligent architectures for flexible and secure resource sharing and content caching in 5G networks. 

Qiu \textit{et al.} \cite{Qiu2019DDQL} provided trust in SDN Industrial IoT networks using Blockchain consensus protocols. In their design, Blockchain collects, synchronizes, and distributes network views between distributed SDN controllers. To improve the throughput, they used a Dueling Deep Q-Learning (DDQL) approach to jointly optimize view changes, access selection, and computational resource allocation. Deep Q-Learning (DQL) was also used in Distributed Software-Defined Vehicular Network (DSDVN) to adapt to the variety of data, network-flow, and vehicle types \cite{Qiu2018DQL}. To increase the throughput of a permissioned Blockchain, consensus schemes were used to reach consensus efficiently and securely in DSDVN using DQL. It optimizes compute and network resources by jointly considering the trust features of Blockchain nodes to improve the throughput.

Blockchain integration with all those technologies was a major factor in the deployment of smart environment applications. Sharma \textit{et al.} \cite{sharma2017block}, for example, used Blockchain to allow for decentralized coordination and control for vehicular networks in smart cities. Moreover, Sharma and Park \cite{SHARMA2018Blockchain} integrated Blockchain and SDN to provide a two-layer network architecture that was specially designed for smart cities. It is composed of core and edge networks, and leverages the benefits of both centralized and distributed architectures. Their design supports IoT heterogeneity and provides a scalable and secure architecture using edge computing. They used a memory-hardened PoW scheme to enforce distributed privacy and security, and to avoid tampering of information by attackers. Sharma \textit{et al.} \cite{Sharma2019Blockchain} used a private version of Ethereum to simulate a distributed framework for automotive industries in smart cities. They proposed a novel miner node selection algorithm to increase trust, provide protection, and save time and cost in automotive supply chain ecosystems that are owned by different organizations.

\section{Open Research Problems}
\label{sec:res}
A lot of work is still needed to allow for smoother integration between Blockchain and IoT technologies to create smarter things. Law and regulation issues are some of the main problems that are discussed in the literature for using Blockchain for basic monetary transactions. Hence, these issues will directly impact machine-to-machine monetary transactions using Blockchain. Akins \textit{et al.} \cite{PittsburghTaxReview} discussed income taxation of cryptocurrency transactions, such as Bitcoins, when used for purchases with monetary value. Sapovadia \cite{SAPOVADIA2015253} discussed the legal issues in cryptocurrencies that are similar to those of foreign currencies. Emelianova and Dementyev \cite{Taxation2020} argued for a unified supranational legal act for cryptocurrencies, similar to the European Union (EU) directives. They discussed the provisions in many European and Asian countries for the use and taxation of cryptocurrencies, which differs from one government to another. Omololu \cite{Omololu2020} emphasized the need for full law enforcement of Blockchain applications as they still do not conform to current legal structures. This requires countries to supervise Blockchain integration in different applications and domains, including IoT, to ensure that they comply with the law.

To create an IoT-oriented Blockchain platform, both hardware and software should be highly optimized to perform Blockchain operations. IBM has taken the lead in this path by creating a 10-cents tiny edge CPU architecture that can efficiently run Blockchain operations \cite{IBMAnchor}. It can be embedded into IoT devices to support Blockchain operations. IBM called this project "Crypto Anchors", as they want to anchor physical objects into Blockchain-IoT applications. However, the difference between such CPU architectures and standard Computer CPUs requires specific Blockchain implementations for Crypto Anchor CPUs. Hence, we believe that this is a good research direction to create an operating system for those CPUs that is capable of bridging this gap, such as the work by Wright and Savanah \cite{wright2019operating}. Building Blockchain-oriented CPUs, firmware, and operating systems is a great step towards more robust Blockchain-IoT integration. However, we believe that the research should also continue on improving Blockchain architectures, including consensus algorithms and communication protocols.

In terms of consensus algorithms, there has been an effort to replace the most secure consensus algorithm in practice, i.e. PoW. The main reason is the power consumption of PoW, which does not fit resource-limited IoT devices. PoS \cite{king2012ppcoin} and DPoS \cite{larimer2014delegated} are two promising alternatives, but they are still criticized for not being as secure as PoW \cite{poelstra2014distributed}. Ethereum plans to migrate from using PoW to PoS \cite{b_i_m_2019}, because it is currently the best alternative for PoW \cite{saleh2020Blockchain}. DPoS is currently adopted in EOS \cite{io2017eos} and few other Blockchain implementations. There is a strong debate around the level of decentralization in DPoS and PoW \cite{li2020comparison}. To show the difference, Li and Palanisamy \cite{li2020comparison} studied a DPoS-based cryptocurrency for social media, called Steem \cite{steemit}, and the PoW-based Bitcoin Blockchain. They showed that Bitcoin is more decentralized compared to Steem among top miners, but less decentralized in general due to Bitcoin mining pools. 

Consensus algorithms are a major factor in determining Blockchain performance, and that is why researchers try to increase the security and lower power consumption of these algorithms \cite{PoWvsBFT, BitcoinBackbone}. Even lightweight Blockchain platforms that are built specifically for resource-limited IoT applications have some drawbacks. DAG-based platforms for example, like IoTA, suffer from double-spending attacks. Hence, there is always a security vs. performance trade-off when choosing between different Blockchain platforms or consensus algorithms. Such trade-off should be carefully selected to meet the different requirements of different IoT applications. Pongnumkul \textit{et al.} \cite{privateBlockchains2017}, for example, studied the trade-off between choosing the most secure PoW vs. the fastest PBFT consensus algorithms in Ethereum and Hyperledger Fabric, respectively.

However, it is sometimes necessary to compare different Blockchain platforms using factors other than consensus algorithms. Developers can falsify Blockchain performance, and attract investors only based on the consensus performance. However, choosing a Blockchain platform based solely on the consensus performance can badly affect its performance in large scale IoT networks or in smart city applications. Hence, Zheng \textit{et al.} \cite{zheng2018Blockchain} argued for Blockchain testing schemes to help practitioners select Blockchain platforms that fit the requirements of different IoT applications. They also discussed the drawbacks of mining pools in public Blockchain implementations, which can cause loss of decentralization. Another Blockchain related problem in IoT applications are Smart Contract bugs. These bugs shorten the life-cycle of that code and loosen its agility in the ever changing IoT networks. 

Furthermore, using Oracles \cite{Oracles}, which are trusted third party information sources, exposes Blockchain to lose its inherited security. Oracles provide external data and information to Blockchain Smart Contracts to enrich the capabilities of Blockchain applications, including IoT applications. Oracles query, verify, and authenticate external data sources to provide trust for such sources. Blockchain should trust oracles, since they make decisions based on the data they provide. However, the trust issues of Oracles will directly impact Blockchain security that was meant to work in trustless environments. Oracles might still suffer from centralization, collusion, Sybil, and Man-in-the-middle attacks \cite{OraclesChall}. They are also possibly exposed to physical attacks on IoT devices that are usually the source for external Blockchain data. An example of physical attacks in IoT food or drug supply chain systems is the displacement of temperature or GPS sensors that are usually attached on supply chain shipping trucks. The displacement of IoT devices feeds Blockchain ecosystems with falsified information, which causes those systems to loose their security features, and hence to fail.

51\% attacks, cost, regulations, confirmation time, forks, and scalability are common technical Blockchain-related problems \cite{lin2017survey}. Scalability, for example, has been discussed a lot in literature, and different solutions have been proposed \cite{Scaling}. In addition, there are also domain-specific challenges, such as the challenges for using Blockchain for AI \cite{Blockchain2018Intrusion}, Security \cite{Blockchain2018Intrusion}, Healthcare \cite{Siyal_2019}, Education \cite{Garcia-Font2020}, Product Traceability \cite{KAMBLE2020101967}, E-Voting \cite{Amr2020Voting, KHAN202013, voting2018Yang, _abuk_2018}, and IoT applications. There is a need to reduce Blockchain energy consumption and operation cost to make it feasible to integrate with various technologies, including IoT. However, security, privacy, scalability, and Oracle inherited Blockchain issues need to be addressed before looking for integration issues \cite{Luo2020}. Finally, real deployment of Blockchain-IoT solutions in smart city prototypes might reveal new issues compared to what simulation results currently demonstrate.

To focus more on Blockchain-IoT integration challenges, Makhdoom \textit{et al.} \cite{MAKHDOOM2019251} used a test case for a supply chain monitoring system. The challenges include the lack of IoT-centric consensus protocols, IoT-based transaction validation rules, IoT-oriented Blockchain interfaces, and storage capacity for IoT data. They added other Blockchain-related challenges, like consensus finality, resistance to DoS attacks, fault tolerance, scalability, and transaction volume. The test case required a secure and synchronized software upgrade scheme for IoT devices and the underlying Blockchain platform. The upgrade scheme can fix bugs and protect the system against new vulnerabilities. In addition, we need to be careful when integrating SDN and Blockchain technologies to fully benefit from SDN features for IoT applications. Blockchain is decentralized by nature, whereas SDN controllers are supposed to be centralized. Moreover, SDN controllers need to update the flow-rule tables to the forwarding devices in real-time, while Blockchain consensus works periodically on a longer time frame.

The challenges brought by integrating Blockchain with Cloud, Fog, and Edge Computing directly relate to IoT-Blockchain challenges. That is because these technologies are mainly meant to support IoT networks. The lack for perfectly implemented Blockchain-specific IoT infrastructures, and the absence of energy-efficient mining are some of those challenges \cite{8543246}. Authentication, adaptability, network security, data integrity, verifiable computation, and low latency are requirements for integrating Blockchain with Cloud, Fog, and specially Edge Computing \cite{8624417}. Yang \textit{et al.} \cite{8624417} identified load balancing, task offloading, resource management and function integration on heterogeneous platforms as challenges to be addressed for successful integration. Addressing all these issues is essential to support next generation applications in fully automated futuristic smart environments. In such environments, all these technologies should be smoothly integrated, flawlessly functioning, and securely handling IoT data.

\section{Discussion and Conclusion}
\label{sec:con}
In this paper, we present the strengths of Blockchain beyond its traditional use for monetary and digital asset trading. We focus on Blockchain integration with IoT to create a futuristic view of smart environments. Implementing such autonomous smart environment architectures will simplify human lives and increase their effectiveness. We discuss the role of AI, SDN, Cloud, Fog, and Edge Computing in enhancing the capabilities of Blockchain-IoT applications, and providing such automation in smart environments. Blockchain augments IoT applications with automation, security, privacy and many other features that are essential for smart environments. We showed in this work how Blockchain was able to address a number of issues, limitations, and challenges in all those technologies.

Table \ref{tab:compareDesg} shows the level of technological integration of some Blockchain-IoT applications and prototypes in the literature. The table also shows the recent research interest to integrate more technologies into such systems. Powered by Blockchain-IoT integration, these prototypes served different needs and solved different problems in different smart applications. To build those systems, some of the authors had to create their own Blockchain implementation and/or its consensus algorithm. Specific application requirements usually ask for different Blockchain characteristics that are not usually available in traditional implementations. Hence, there is a need for a Blockchain implementation with the possibility to plug in new features when needed by developers and practitioners. This will allow the industry and the research community to focus more on developing more smart applications, and mitigating different technological limitations.

\begin{table*}
  \caption{Technological integration of Cloud, Edge, Fog, SDN, and AI technologies into Blockchain-IoT prototypes.}
  \label{tab:compareDesg}
  \setlength{\tabcolsep}{3pt}
    \begin{tabular}{|l|c| *{4}c | c|c|c|c|c|}
        \hline
        \textbf{Paper} & \textbf{Year} & \multicolumn{4}{c|}{\textbf{Integration with}} &
        \textbf{Platform}&\textbf{Functionality}&\textbf{Mining}&\textbf{Type}&\textbf{Transactions}\\[2ex] 
        & & \rot{Cloud} & \rot{Edge/Fog} & \rot{SDN} & \rot{AI} &&&&& \\\hline\hline
        
        \cite{sensor2014money} & 2014 &&&&  &Bitcoin&Data Market&PoW&Public&UTXOs \\\hline 

        \cite{Zhang2015IoT}\textsuperscript{+} & 2015 &\OK&&&  &Bitcoin&Data Market&PoW&Public&UTXOs \& Scripts \\\hline 
        
        \cite{DecPriv} & 2015 &&&&  &Bitcoin\textsuperscript{6}&Decentralized Data Management&PoW&Public&Data \& Queries \\\hline
        
        \cite{bahga2016blockchain} & 2016 &\OK&&&  &Ethereum&D2D Communication \& Marketplace&PoW&Private&Smart Contracts \\\hline 
        
        \cite{dorri2016blockchain}\textsuperscript{+} & 2016 &\OK&&&  &Bitcoin\textsuperscript{6}&Smart Homes&&Public&IoT Data \\\hline  

        \cite{axon2016a} & 2017 &&&&  &NameCoin&Privacy-Aware PKI&PoW&Public&Digital Signatures \\\hline 
        
        \cite{Dorri2017SmartHome}\textsuperscript{+} & 2017 &\OK&&&  &Bitcoin\textsuperscript{6}&Smart Homes&&Private&IoT Data \\\hline 
        
        \cite{Huh2017Managing} & 2017 &&&&  &Ethereum&D2D Communication&PoW&Public&Smart Contracts \\\hline 
        
        \cite{UAV2017Kapitonov} & 2017 &&&&\OK  &Ethereum&D2D Communication&PoW&Public&Smart Contracts \\\hline     

        \cite{Sharma2017DistBlockNet}\textsuperscript{+} & 2017 &&&\OK&  &&Decentralized SDN (DSDN)&PoW&& \\\hline 

        \cite{SIKORSKI2017234} & 2017 &&&&  &MultiChain&D2D Marketplace&Round Robin&Private&Digital Assets \\\hline
        
        \cite{Sharma2017IoT}\textsuperscript{+} & 2018 &\OK&\OK&\OK&  &&Distributed Cloud&PoSer\textsuperscript{3}&Private&Smart Contracts\textsuperscript{1} \\\hline 
        
        \cite{Almadhoun2018Authentication} & 2018 &\OK&\OK&&  &Ethereum&Authentication&PoW&Public&Smart Contracts \\\hline 

        \cite{Liu2018Offloading} & 2018 &&\OK&&  &Wireless BC\textsuperscript{7}&Offloading PoW Computations&PoW&& \\\hline  
        
        \cite{PoWR} & 2018 &\OK&&&  &Bitcoin\textsuperscript{6}&Wireless Network Virtualization&PoWR\textsuperscript{5}&Public&Network Data \\\hline

        \cite{Qiu2018DQL} & 2018 &&\OK&\OK&\OK  &Fabric&Vehicular Networks&PBFT&Private&Smart Contracts\textsuperscript{1} \\\hline  
        
        \cite{Sharma2018SDN}\textsuperscript{+} & 2018 &\OK&\OK&\OK&  &&DSDN for Edge Computing&PoW&& \\\hline 

        \cite{SHARMA2018Blockchain}\textsuperscript{+} & 2018 &&\OK&\OK&  &Ethereum&Smart Cities&PoW&Private&IoT Data Hashes \\\hline 
        
        \cite{Xiong2018MobileBlockchain} & 2018 &\OK&\OK&&  &Ethereum&Edge Computing Mining&PoW&Private&IoT Data \\\hline 
        
        \cite{BlockchainRL2019}\textsuperscript{+} & 2019 &\OK&\OK&&\OK  &&Intelligent Wireless Networks&PBFT&Consortium&Network Data \\\hline 
        
        \cite{Qiu2019DDQL} & 2019 &&\OK&\OK&\OK  &&DSDN Consensus&PBFT&Private&Traffic/Control Data \\\hline  

        \cite{fusion2019} & 2019 &&\OK&\OK&  &Bitcoin\textsuperscript{6}&Wireless Network Virtualization&&Public&Network Data \\\hline 

        \cite{Sharma2019Blockchain} & 2019 &&&&  &Ethereum&Supply Chain Management&PoW&Private&Smart Contracts\textsuperscript{1} \\\hline 

        \cite{Xiong2019CFC} & 2019 &\OK&\OK&&  &Ethereum&Offloading PoW Computations&PoW&Private&Random Data \\\hline 
        
        \cite{TULI201922} & 2019 &\OK&\OK&&  &Bitcoin\textsuperscript{6}&Data Privacy and Integrity&PoW&&IoT Data \\\hline 
        
        \cite{Huo2020} & 2020 &\OK&\OK&\OK&  &&Securing Traffic Measurements&&&Network Statistics \\\hline 
        
        \cite{Medhane2020EdgeCloudSDN}\textsuperscript{+} & 2020 &\OK&\OK&\OK&  &&Distributed Network Security&&&IoT \& Network Data \\\hline 

        \cite{WILCZYNSKI2020Modelling} & 2020 &\OK&&&  &Their Own&Secure Cloud Scheduler&PoSch\textsuperscript{2}&Public&Task Scheduling Info \\\hline 

        \cite{Wu2020Cooperative} & 2020 &\OK&\OK&\OK&  &Their Own&Secure Fog Computing&Cooperative&&Access Control Lists \\\hline 
        
        \cite{9025213} & 2020 &&&\OK&\OK  &Bitcoin\textsuperscript{6}&DSDN Consensus&BFT\textsuperscript{4}&Private&Traffic/Control Data \\\hline      
        
        \hline
        \multicolumn{11}{l}{
            \textsuperscript{1}Contract execution beside monetary and data transactions.
            \textsuperscript{2}Proof-of-Schedule. 
            \textsuperscript{3}Proof-of-Service. 
            \textsuperscript{4}Aardvark, RBFT, \& PBFT.
            \textsuperscript{5}Proof-of-Wireless-Resources.
        }\\
        \multicolumn{11}{l}{
            \textsuperscript{6}A different Bitcoin implementation based on the same concepts and architecture.
            \textsuperscript{7}A simulation for a wireless Blockchain implementation.
        }\\
        \multicolumn{11}{l}{
            \textsuperscript{+}The design or prototype has a hierarchical or multilayer architecture. \textbf{Blank Cells:} Information was not provided in the corresponding paper. 
        }

    \end{tabular}
\end{table*}

Figure \ref{fig:smart} shows a simplified architecture for smart environments using the technologies discussed in this paper. It shows the use of Blockchain to securely share and store IoT data in trustless environments. In this architecture, Blockchain can be deployed using Cloud, Fog, or Edge resources, or even a combination of them. The AI-powered SDN traffic control can manage the traffic flow of IoT data in a dynamic smart way. AI can also be used to provide best pricing for IoT data that needs to be sold in such a system, like sensor data. It can be also used to provide the best target to process this data using the Cloud, Fog, or Edge computing. In addition, private data can be securely stored in Blockchain, and can be obtained by authentic users using different encryption and security measures. Blockchain will work as an access control mechanism for data access, including IoT data, for IoT devices and their users.

\Figure[t!](topskip=0pt, botskip=0pt, midskip=0pt)[width=0.48\textwidth]{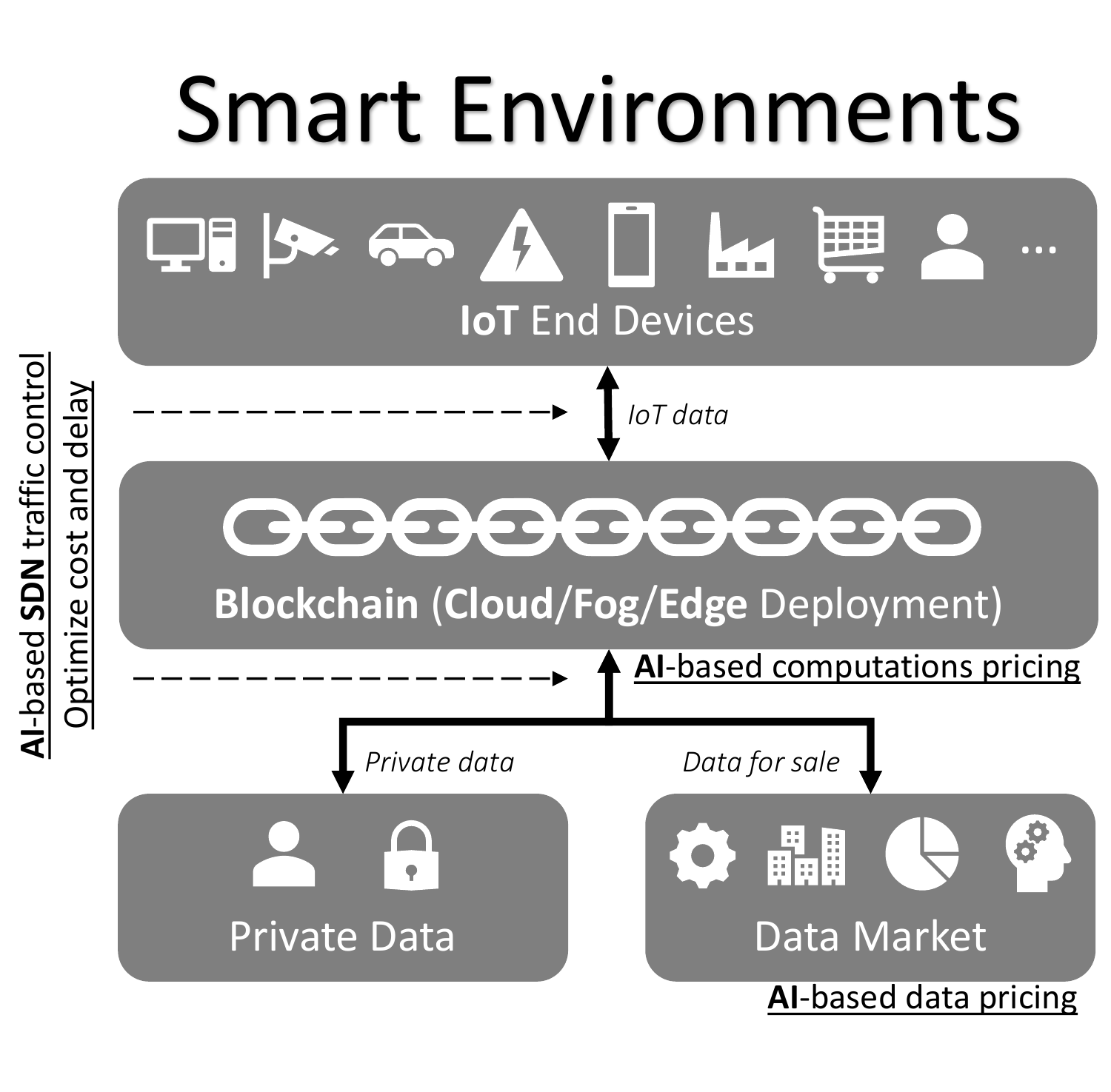}
{Architecture of Smart Envrinoments.\label{fig:smart}}

There are a number of challenges and open research problems that still need to be tackled to provide a smoother technological integration. A major problem of such integration is the difficulty of testing using physical deployment in real smart environments in order to reveal hidden issues. Physical deployment is needed because simulation results alone are insufficient to demonstrate the performance and issues in such systems. In addition, creating a general purpose Blockchain platform that can be easily adapted to solve different problems is of a great importance for the research community. Creating such a general purpose platform will remove the burden that is imposed by modifying Blockchain architectures to meet certain features and requirements.


\bibliographystyle{IEEEtran}
\bibliography{access.bib} 

\begin{IEEEbiography}
[{\includegraphics[width=1in,height=1.25in,clip,keepaspectratio]{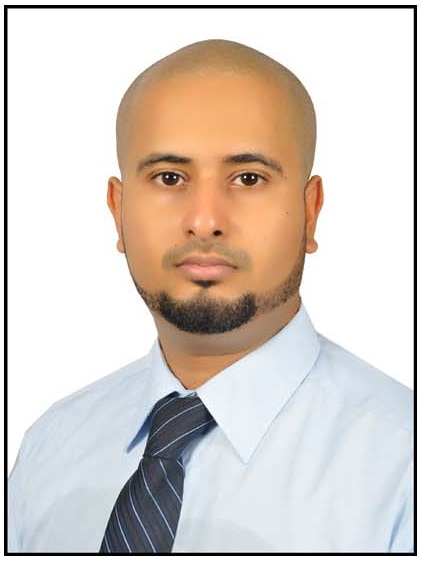}}]{Maad Ebrahim} is currently a Ph.D. student at the Department of Computer Science and Operations Research (DIRO), University of Montreal, Canada. He received his M.Sc. degree in 2019 from the Computer Science Department, Faculty of Computer and Information Technology, Jordan University of Science and Technology, Jordan. His B.Sc. degree in Computer Science and Engineering has been received from the University of Aden, Yemen, in 2013. His research experience includes Computer Vision, Artificial Intelligence, Machine learning, Deep Learning, Data Mining, and Data Analysis. His current research interests include Fog and Edge Computing technologies, Blockchains, and Reinforcement Learning.
\end{IEEEbiography}

\begin{IEEEbiography}
[{\includegraphics[width=1in,height=1.25in,clip,keepaspectratio]{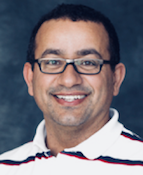}}]{Abdelhakim Hafid} spent several years as the Senior Research Scientist with Bell Communications Research (Bellcore), NJ, USA, working in the context of major research projects on the management of next generation networks. He was also an Assistant Professor with Western University (WU), Canada, the Research Director of Advance Communication Engineering Center (venture established by WU, Bell Canada, and Bay Networks), Canada, a Researcher with CRIM, Canada, the Visiting Scientist with GMD-Fokus, Germany, and a Visiting Professor with the University of Evry, France. He is currently a Full Professor with the University of Montreal. He is also the Founding Director of the Network Research Laboratory and Montreal Blockchain Laboratory. He is a Research Fellow with CIRRELT, Montreal, Canada. He has extensive academic and industrial research experience in the area of the management and design of next generation networks. His current research interests include the IoT, fog/edge computing, blockchain, and intelligent transport systems.
\end{IEEEbiography}

\begin{IEEEbiography}
[{\includegraphics[width=1in,height=1.25in,clip,keepaspectratio]{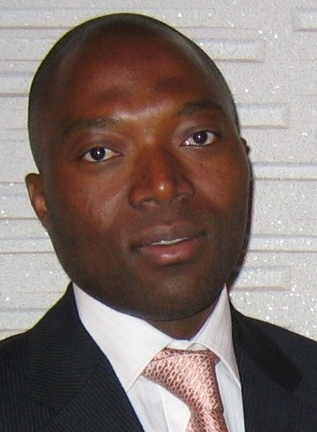}}]{Etienne Elie} is Solutions and Systems Architect and Engineering Lead at Intel Corporation, California, USA. Prior to joining Intel Corporation, Dr. Elie was the technology and engineering manager for CARTaGENE, a public research platform and biobank of the Sainte-Justine Learning Hospital. He also served as ASIC Architecture Engineer at Nortel Networks and Advanced Micro Devices (AMD). Before moving to the US, Elie spent a short period of time with PSP Investments, one of Canada’s largest pension investment managers. Beside his role at Intel Corporation, Dr. Elie is a key contributor for the development of a large-scale general-purpose neuromorphic Community Infrastructure (CI). Dr. Elie holds a Ph.D. in Computer Architecture from Université de Montréal, with focus on optimization of data movements in computer systems. He also holds a master's degree, and Bachelor of Science in Engineering with great distinction.
\end{IEEEbiography}

\EOD
\end{document}